# Dissolving yourself in connection to others: Shared experiences of ego attenuation and connectedness during group VR experiences can be comparable to psychedelics

David R. Glowacki,[1,2,*] Rhoslyn Roebuck Williams,[1,3,4] Mark D. Wonnacott,[3,4] Olivia M. Maynard,[1,5] Rachel Freire,[1,6] James E. Pike,[1,2] Mike Chatziapostolou[1,2]

[1]Isness-D Node Network; [2]ArtSci International Foundation, Bristol, UK; [3]Intangible Realities Laboratory, University of Bristol, Bristol, UK; [4]Centre for Computational Chemistry, University of Bristol, Bristol, UK; [5]School of Psychological Science, University of Bristol, UK; [6]Rachel Freire Studio, London, UK

With a growing body of research highlighting the therapeutic potential of experiential phenomenology which diminishes egoic identity and increases one's sense of connectedness, there is significant interest in how to elicit such 'self-transcendent experiences' (STEs) in laboratory contexts. Psychedelic drugs (YDs) have proven particularly effective in this respect, producing subjective phenomenology which reliably elicits intense STEs. With virtual reality (VR) emerging as a powerful tool for constructing new perceptual environments, we describe a VR framework called 'Isness-distributed' (Isness-D) which harnesses the unique affordances of distributed multi-person VR to blur conventional self-other boundaries. Within Isness-D, groups of participants co-habit a shared virtual space, collectively experiencing their bodies as luminous energetic essences with diffuse spatial boundaries. It enables moments of 'energetic coalescence', a new class of embodied phenomenological intersubjective experience where bodies can fluidly merge, enabling participants to have an experience of including multiple others within their self-representation. To evaluate Isness-D, we adopted a citizen science approach, coordinating an international network of Isness-D 'nodes'. We analyzed the results (N = 58) using 4 different self-report scales previously applied to analyze subjective YD phenomenology (the inclusion of community in self scale, ego-dissolution inventory, communitas scale, and the MEQ30 mystical experience questionnaire). Despite the complexities associated with a distributed experiment like this, the Isness-D scores on all 4 scales were statistically indistinguishable from recently published YD studies, demonstrating that distributed VR can be used to design intersubjective STEs where people dissolve their sense of self in the connection to others.

# 1 INTRODUCTION

Describing the sensations that arose immediately following a left-hemisphere stroke, brain scientist Jill Bolte Taylor recounted: *I could no longer define the boundaries of my body. I can't define where I begin and where I end, because the atoms and molecules of my arm blend with the atoms and molecules of the wall, and all I could detect was this energy... I was immediately captivated by the magnificence of the energy around me. And because I could no longer identify the boundaries of my body, I felt enormous and expansive. I felt at one with all the energy that was, and it was beautiful.* (*1*) Recounting the sensations that arose after seeing the Earth from space, astronaut Edgar Mitchell said: *And suddenly I realized that the molecules of my body, the spacecraft, the body of my partners, were all manufactured in some ancient generation of stars. I felt an overwhelming sense of oneness, or connectedness... an insight, an epiphany.* (*2*) During a particularly potent meditation session Zen master Sokei-an Sasaki recounted how: *I lost the boundary of my physical body. I had my skin, of course, but I felt I was standing in the center of the cosmos... I saw people coming towards me, but all were... myself!* (*3*) Recounting the sensations which arose from a high dose of the psychedelic drug (YD) psilocybin, a participant in a 2008 Griffiths et al. study recounted a *feeling of no boundaries, where I didn't know where I ended and where my surroundings began. Somehow I was able to comprehend what oneness is.* (*4*)

These accounts, which express a profound sense of unity with other beings and other objects, stand in contrast to our typical day-to-day perspective, in which we often default to conceptual representations of ourselves (and others) as separate objects rather than coupled interdependent subjects. Yaden et al. have coined the term 'self-transcendent experiences' (STEs) to describe transient mental states in which *the subjective sense of one's self as an isolated entity can temporarily fade into an experience of unity with other people or one's surroundings, involving the dissolution of boundaries between the sense of self and 'other'* (*5*). Research across psychology, neuroscience, philosophy, pharmacology, and theology has drawn attention to the meaning and insight attributed to STEs by those who undergo them. Because they have been studied across so many different knowledge domains, they are described using different terminology, conceptual frameworks, and theoretical lenses, which makes it complicated to perform a comprehensive analysis of their subjective qualities, behavioral effects, and therapeutic potential. In an attempt to draw conceptual links across these various domains, Yaden et al. have proposed the following mental states as STEs: 1) Flow states, (*6*) which arise from focused absorption in an interesting and challenging task; 2) Mindfulness states, (*7*) which are characterized by moment-to-moment, open, nonjudgmental awareness; 3) Awe, (*8*) which often results from the perception of vastness (e.g., immensity in size, number, scope, complexity, etc.), and the subsequent need to accommodate this vastness; 4) Peak Experiences, (*9*) relatively rare events that were originally named by Maslow, and which typically include a feeling of merging with the universe; and 5) Mystical-Type Experiences, (*10*) a particularly intense class of experience often characterized by ineffability, a feeling of internal and external unity, changes in the experience of space and time, and interwoven senses of connectedness, sacredness, and noetic qualities.

The phenomenology associated with different classes of STEs varies in saliency and visceral potency. As such, STEs have a broad range of intensities, some of which are more or less memorable, and thereby more significant for the person undergoing the experience. For example, "losing oneself" whilst reading a book or writing computer code may represent a relatively weak STE; whereas the deep sense of interdependence with the cosmos that arises from a spiritual 'God-encounter' experience represents a considerably more intense form of STE. (*11*) Analysis by Yaden et al. suggests that STEs involve at least two interrelated phenomenological ingredients: 1) an 'annihilation' of the sense of self, accomplished by dissolving self-boundaries and self-salience; and 2) a 'relational' sense of unity with something beyond the self (e.g., others or the natural world). The intensity of an STE is associated with the degree to which the participant experiences these interrelated phenomenological ingredients. While the precise sense of 'self' is difficult to define here, several different studies have nevertheless demonstrated the effectiveness of the so-called Inclusion of Other in the Self (IOS) Scale (*12*) for assessing the degree to which an individual feels a sense of connectedness to others or to their environment. As shown in Fig 1, this scale represents individuals and others as distinct circles.

During weak unitive experiences, overlap between the circles is small; during particularly intense unitive experiences the overlap is more significant, representing an individual's sense that 'the other' forms a significant part of the self.

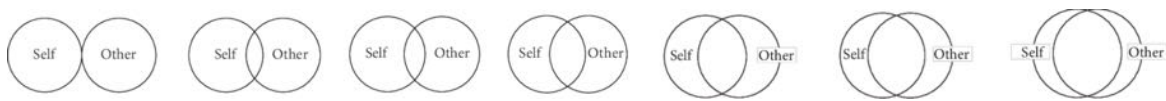

Figure 1: The Inclusion of Other in the Self Scale, which offers a continuum for expressing the sense of unitive feelings

William James suggested that particularly intense classes of STEs – namely 'mystical type experiences' (MTEs), where participants often describe a profound sense of unity with their surroundings – can have lasting positive, transformative, and life-changing effects. Recent analysis of questionnaire data from more than 4000 participants by Griffiths et al. (*11*) seems aligned with James' supposition: individuals who had undergone intense MTEs arising from subjective spiritual experiences rated them to be amongst their most meaningful and spiritually significant life experiences, associated with persistent positive changes in life satisfaction and purpose. Given their long-term therapeutic benefit, increasing research effort has been directed to explore technologies for reliably eliciting MTEs in a laboratory context. In this respect, psychedelic drugs (YDs) have emerged as a particularly promising way of producing subjective phenomenology similar to non-drug MTEs, (*11*) and are associated with lasting therapeutic benefit in addressing depression, addiction, and end-of-life-anxiety. (*4, 13-16*) Unitive experiences of interconnectedness are common during YD experiences, where they are often associated with subjective perception of ego dissolution (*17-19*), near-death-like experiences, (*20*) altered perception of both time and space, (*21*) and a sense of the ineffable. (*11*) For example, a 2006 Griffiths et al. study showed that 67% of participants who had subjective MTEs while taking psilocybin as part of a 'psychedelic psychotherapy' program considered the experience to be amongst the most meaningful experiences of their lives. Despite their potential, YDs face a number of practical challenges to widespread administration: they produce intense phenomenology that can lead to transient fear or panic, (*13*) and therefore require expensive psychotherapy support; YDs also invoke short-term physiological responses like increased heart rate and blood pressure, (*22*) sometimes leading to headaches, nausea, and vomiting; and ongoing questions remain how to determine a YD dose that reliably elicits therapeutic effects but minimizes the aforementioned risks. Finally, YDs face significant regulatory challenges, which vary considerably across cultures.

Given the various challenges associated with YD administration, there has been emerging research exploring non-drug technologies for reliably eliciting STEs. Virtual Reality (VR) has emerged as a particularly interesting candidate (*23, 24*) given its ability to create strong alterations in perceptual phenomenology. For example, Glowacki et al. (*25*) recently described a multi-person VR experience called 'Isness', where four participants co-located in the same physical and virtual space (Fig 2A) experience the collective emergence, fluctuation, and dissipation of their bodies (and those of their co-participants) as diffuse energetic essences. The subjective MTEs produced by this co-located version of Isness (Isness-C) were evaluated (N = 57) using the MEQ30, (*26, 27*) a commonly used YD experience questionnaire. (*28*) Glowacki et al. compared the Isness-C MEQ30 results to those obtained from more than 540 participants in 26 previous YD studies and showed that the Isness-C scores were statistically indistinguishable from MEQ30 scores following moderate-to-high doses of YDs. Isness-C participants reported positive emotions, a unitive sense of connectedness, a weakening of ego boundaries, and awareness of pure presence that was less distracted by ego. Isness-C offers a case study demonstrating that multi-person VR can be used to design group experiences that create the conditions for intense STEs from which participants derive insight and meaning. To distinguish technologies like Isness from psychedelic technologies, Glowacki et al. coined the term *numadelic*. Combining the Greek words *pneuma* (πνευμα, 'breath', 'spirit', or 'soul') and *delein* (δηλειν, 'to reveal', 'to make visible', or 'to manifest'), *numadelic* may be translated as 'spirit-manifesting' or 'spirit-revealing'. The *numadelic* aspect of Isness-C arises from the fact that participant bodies are represented as luminous energetic essences, an aesthetic representation associated with 'spirit' in a variety of wisdom and meditation traditions.

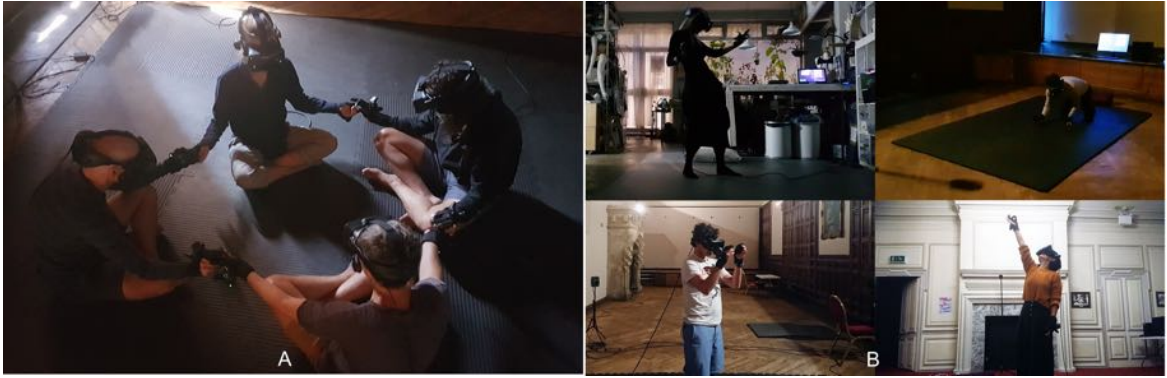

Figure 2: (A) the co-located version of Isness (Isness-C), an intersubjective multi-person group VR experience which Glowacki et al. showed produces MEQ30 scores comparable to moderate to high doses of YDs; (B) the distributed version of Isness (Isness-D), with four participants simultaneously joining a session from different Isness-D nodes distributed across the world.

While Isness-C was (to the best of our knowledge) the first study using the MEQ30 to analyse subjective participant experience in a VR context, previous work has investigated the ability of both 2d and immersive video content to elicit two phenomenological ingredients known to be important components of a subjective YD-MTE, namely: 1) altered visual perception, (*29*) and 2) a sense of awe. (*23, 30*) The former has been studied by Suzuki et al., who presented individuals with immersive 360 videos derived from Google's deep dream convolutional neural nets, rendered in a VR headset. (*31, 32*) Their results (N = 12) suggested that it is possible to induce visual phenomenology similar to psilocybin; however, they were unable to evoke in participants the corresponding sense of temporal distortion that psilocybin produces. The latter – awe – has been the topic of several studies. Broadly, these studies have shown that awe can be reliably elicited by presenting individuals with video content of various environments, e.g.: views from a mountaintop, within a vast open space, seeing earth from space, submerged in the ocean, amongst very tall trees, etc. (*33-38*) Recent fMRI studies suggest that participants presented with awe-eliciting 2d videos have reduced activity in the brain's default mode network, (*39*) which is an ego-reaffirming neurological network that YDs also attenuate. (*15*) Chirico et al. showed that immersive videos rendered in head-mounted-displays (HMDs) are able to elicit a stronger sense of awe compared to standard 2d videos, (*33*) given that VR enables a heightened sense of immersion. (*40, 41*)

The two items which Yaden et al. identified as primary phenomenological ingredients of STEs – 1) a weakened sense of self-boundaries and diminished self-salience; and 2) a 'relational' sense of unity with something beyond the self – are not easy to untangle. The difficulty becomes apparent by undertaking a simple thought experiment, imagining the dynamics that arise when two drops of liquid water coalesce: their individual identities and associated properties fade away as their respective boundaries co-mingle and eventually dissolve into one another, eventually creating a drop with new physical properties. Characterizing the softness, fluidity, or porosity of a body's boundaries is impossible without corresponding observations of the dynamics that unfold when it undergoes some form of relational interaction with another body. The tight relational co-dependence of these phenomenological ingredients, in which one is required to characterize the other and vice-versa, suggests that they are perhaps *best realized in a multi-body, relational context*. This observation highlights a potentially significant gap in YD laboratory studies and efforts to design STEs using VR: *nearly all of the work described to date involves individual, intrasubjective experiences*.

## 2 PRESENT WORK

In what follows, we describe our efforts designing intersubjective group STEs which enable participants distributed across the world (e.g., as shown in Fig 2B and 3A) to collectively co-habit a shared virtual space and undergo an embodied phenomenological experience of the overlap shown in the Inclusion of Other in the Self Scale in Fig 1 (albeit in 4d ($x, y, z, t$) rather than 2d). Specifically, we constructed a multi-person cloud-mounted VR experience called Isness-distributed (Isness-D), which is illustrated in Fig 3A and Fig 3B. Harnessing the unique affordances of distributed multi-person VR, Isness-D's *numadelic aesthetic* blurs conventional boundaries between self and other. Isness-D participants are rendered as luminous energetic essences with radiance concentrated at the heart centre. Compared to the hard boundaries illustrated in the IOS Scale (Fig 1), Isness-D body boundaries are diffuse, fuzzy, and soft: they extend beyond the limits of the physical body, making it difficult to specify clearly where one body ends and another begins, as shown in Fig 3B. This energetic aesthetic takes inspiration from subjective 1st person accounts describing intense unitive experiences of the sort described at the beginning of this article, which often report a heightened sensitivity to a more fundamental energetic essence, and to the molecular and atomic constituents that combine to create our bodies and everything around us. The *numadelic* aesthetic is aligned with fundamental insights of modern physics (e.g., quantum mechanics), which emphasizes that objects have wave-like energetic essences and behaviors. Unlike their classical counterparts, waves are energetic processes and fundamentally non-local. This emphasis on matter's *energetic* essence (rather than a localized materiality) is famously encapsulated by David Bohm's description of matter as 'frozen light'. Karl Popper similarly observed: *Matter turns out to be highly packed energy, transformable into other types of energy; and therefore something in the nature of a process… the results of modern physics suggest that we should give up the idea of a substance or essence… there is no self-identical entity persisting during all changes in time…* (*42*)

The embodied experience of energetic overlap, illustrated in Fig 3A and 3B – where participants can fluidly merge their energetic virtual bodies with those of the others in the group – we refer to as 'energetic coalescence'. To the best of our knowledge, 'energetic coalescence' *represents a new class of phenomenological experience which is only possible within a multi-person distributed VR environment – i.e., it cannot be realized in any other way*. This distinguishes Isness-D from previous work using VR to investigate STEs, which use VR to *simulate* STEs that are in principle available in the real world (YD visuals in the case of Suzuki, or awe experiences in the case of Chirico). In this respect, *'energetic coalescence'* is similar to the 'body-transfer' work of Slater and co-workers (which enables a VR participant to inhabit the body of another person) – i.e., *a 'body-transfer' experience is only possible* using VR. (*43*) Whereas 'body-transfer' is an individual experience, 'energetic coalescence' is a group experience. The 'energetic coalescence' experience represents a particularly explicit form of 'identity fusion', which 'entails a visceral feeling of oneness with the group'. (*44*) As an experience of intense togetherness that temporarily transcends social structures, dissolving the norms that typically characterize relationships within structured institutional contexts, 'energetic coalescence' represents an example of what Victor Turner called 'communitas'. (*45*)

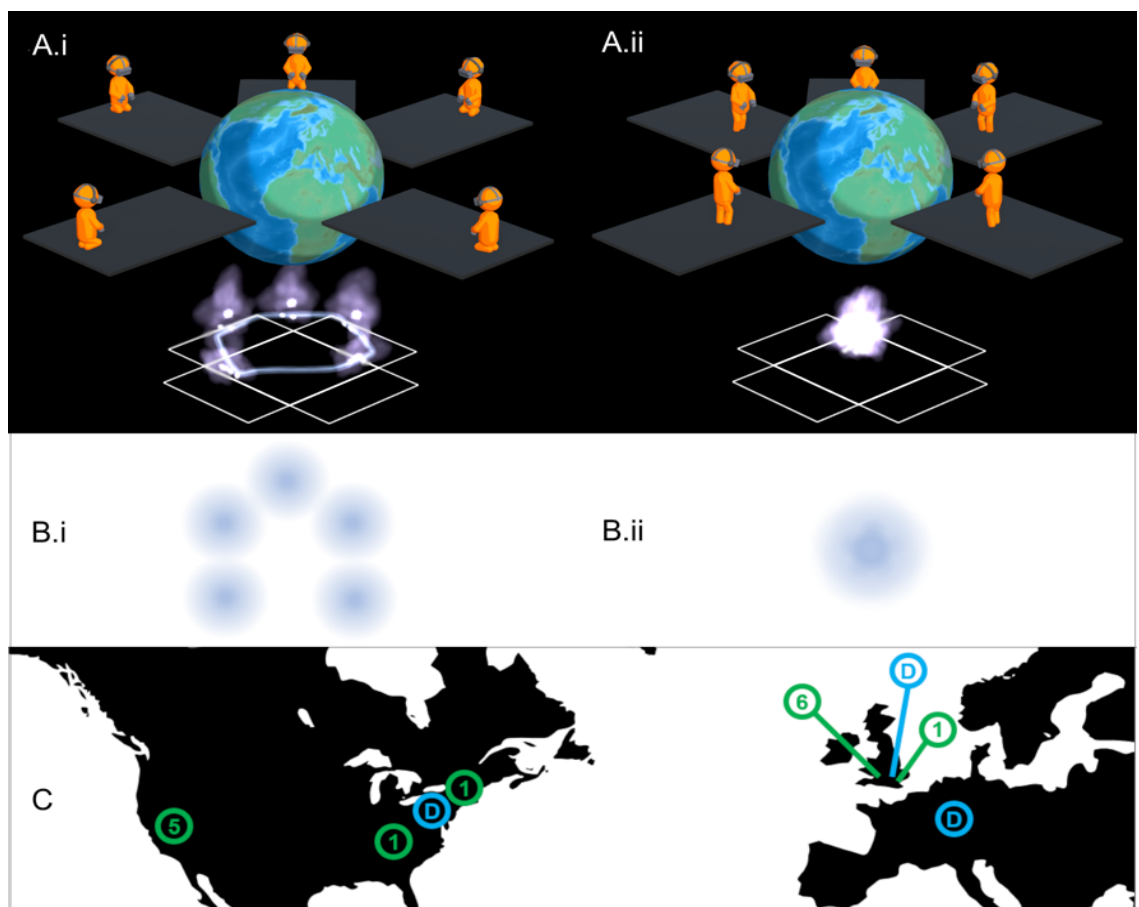

Figure 3: (A.i) shows 4 participants + 1 facilitator joining Isness from various 'nodes' distributed across the world. Participants are kneeling at the edge of their space and are represented in-world as energetic essences connected by a tangible dynamical molecular thread (which they 'hold' in their hands). (A.ii) shows participants undertaking 'energetic coalescence'. Standing at the center of their respective mats, their energetic bodies overlap as they occupy the same position in the shared virtual space. The pattern on the floor in (A.i) and (A.ii) shows how participants' local 'play spaces' were oriented within the virtual space. (B.i) & (B.ii) show IOS diagrams corresponding to (A.i) and (A.ii), for analogy with Fig 1. (C) shows the international Isness 'node' locations as circular markers labelled with the total number of nodes in that location. There were 5 nodes in the San Francisco Bay Area (USA), 1 node in Tennessee (USA), 1 node in Massachusetts (USA), 6 nodes in Bristol (UK), and 1 node in London (UK). The Isness server was mounted on data centers D in London (UK), Frankfurt (Germany) and Washington D.C (USA).

Isness-D is a group VR experience in which participants experience the emergence, dissipation, and fluctuation of their own bodies and the body of a simulated molecular 'organism' as energetic essences. Participants are guided by a combination of a live in-world facilitator and a pre-recorded narrative (further details in sections 3.7 – 3.9). It grew out of the open-source Narupa project, (*46*) a multi-person room-scale VR framework (*47*) we initially designed to enable individuals (*46, 48, 49*) and then groups (*46, 50-52*) to cohabit virtual environments where they can reach out and sculpt the dynamics of simulated molecular objects, made possible by mounting the simulation on GPU-accelerated cloud servers. (*46, 48, 49*) Our work over the years suggests that people interacting with real-time molecular simulations (who from the outside appear to be doing little more than waving their hands around in the air) are able to 'feel' differences in the simulated dynamical flexibility of different molecular objects. (*46, 53*) This highlights a crucial affordance of VR: *a modality that enables a tangible sensory experience of touching phenomena whose essences* *are purely energetic*. For designing STEs, this is a particularly useful affordance, given that it is aligned with subjective YD accounts where participants report sensitivity to the energetic nature of their environment – famously encapsulated by Aldous Huxley's account of his mescaline experience. In a vase of flowers, he reported seeing "*the miracle, moment by moment, of naked existence… flowers shining with their own inner light and all but quivering under the pressure of the significance with which they were charged… [even] the folds of my grey flannel trousers were charged with is-ness.*" (*54*) The presence of a molecular object whose dynamics were simulated in real-time was therefore an important intersubjective aspect of Isness-D. It represented for participants a kind of tangible, fluctuating organism (referred to at various points in the narrative as a 'molecular organism' or 'energetic thread') with a purely energetic essence, inviting people to recognize that their own essential energetic quality is shared with every other object (living and non-living) in the natural world. The depth of the Isness-D matter-energy narrative was reinforced by the fact that the dynamics of the molecular object were calculated in real-time using a state-of-the-art GPU-accelerated computational biophysics engine. (*55*) This sophistication anchored the Isness-D narrative in physical and scientific reality, encouraging participants to reflect on the fact that everyday material objects *are actually* constructed from the dynamical choreography of molecular organisms whose essences are fundamentally energetic, (*49*) an insight which arises in many first-person STE accounts (e.g., those at the beginning of this article). As shown in Fig 3A.i, *multiple participants rendered as energy bodies, connected across different locales and time zones via an energetic molecular thread, represents a class of phenomenological experience that is only possible using the affordances of multi-person distributed VR.*

Given that our laboratory (like many labs) was shut down as a result of COVID restrictions, we evaluated Isness-D using a 'citizen science' approach, coordinating a set of 14 "Isness-nodes" distributed across the world (Fig 3C). Each node operated as a portal where a single participant could virtually enter into the shared Isness-D experience along with participants from other nodes. To analyse Isness-D, we again utilized the MEQ30 in order to facilitate direct comparisons with Isness-C. To assess other phenomenological ingredients known to be important during intense STEs, we used three additional measurement scales. First, we measured the extent to which participants in Isness-D experienced a unitive sense of connectedness with the various participants joining from other Isness-D nodes. To measure this, we used an adapted version of the Inclusion of Community in Self (ICS) Scale, which is essentially a 5-item version of the IOS scale shown in Fig 1, where participants were asked to rate the extent to which a circle representing the community of 'Other Isness Participants' overlapped with their sense of self. (*56*) Second, we measured the extent to which participants experienced a diminished sense of egoic identity, which (as discussed above) is closely interrelated with the intensity of the unitive experience, prominent within MTEs and YD experiences. To do this, we utilized the ego-dissolution inventory (EDI), which measures aspects of both ego dissolution and ego inflation, and has been utilized in prior YD studies. (*57-60*) Third, given that Isness-D was constructed as an intersubjective group experience, we attempted to measure the extent of 'communitas', which has been defined as an experience of intense togetherness and shared humanity that temporarily transcends social structures, mediated by an anti-structural and often ritualized 'liminal' phase of equality amongst community members. (*45*) To do this, we used the recently validated psychedelic communitas scale from Kettner et al., (*61*) who showed that communitas experienced during intersubjective group YD experiences predicts enduring changes in psychological well-being and connectedness, with associated improvements in wellbeing, depressive symptoms, trait anxiety, and interpersonal tolerance.

Isness-D offers an interesting case study in developing technologies, methods, strategies, and protocols to carry out distributed multi-person studies in scenarios where face-to-face laboratory studies and clinical interactions are difficult. The lockdowns, social distancing, and uncertainty that has arisen as a result of the public response to the COVID-19 pandemic have created the conditions for increased social isolation and loneliness, which are correlated with feelings of fear, anxiety, stress, depressive thoughts, and helplessness. (*62-64*) While the longer term effects of COVID remain to be determined, past studies have highlighted social disconnectedness to be a key feature of depression, (*65*) while feelings of connectedness are associated with reduced depressive symptoms. (*66*) By encouraging participants to imagine their own bodies and those of others as energetic essences whose luminosity extends beyond the boundaries of their material form, Isness-D offers participants new experiences of connectedness, which may help address the isolation resulting from COVID. In this respect, experiences like Isness-D respond to recent calls to use human-computer-interaction (HCI) technologies to design meaningful experiences (*67*) which deal with the fundamentals of human existence (e.g., mortality, identity, isolation, meaning), (*68, 69*) in order to stimulate alternative narratives and visions within interconnected systems (*70, 71*) and explore new forms of spiritual sensitivity. (*72, 73*)

## 3 METHODS

### 3.1 Ethics

Ethics approval for the study described herein was obtained from the Faculty of Science Research Ethics Committee at the University of Bristol (ethics approval code: 111003). Participants read an online study information sheet and completed a tick box consent form before signing up to an Isness-D session. The study was conducted according to the revised Declaration of Helsinki (2013) and the 1996 ICH Guidelines for Good Clinical Practice E6(R1). All participants were at least 18 years old, and consented to their data being gathered and published. To minimize participant risk, we adopted VR guidelines in line with those recommended by Madary and Metzinger. (*74*)

### 3.2 Software & Hardware

Isness-D is an approximately 50-minute experience with conceptual roots in earlier work by Glowacki and co-workers. (*25, 75-77*) As a fork of the Narupa project, Isness-D is designed around a client/server architecture enabling each VR client access to the positional data of all the other participants, and a shared real-time molecular simulation of a 40-Alanine peptide macrocycle (40-ALA) whose dynamics are calculated in real-time using OpenMM, a GPU-accelerated computational biophysics engine. (*55*) Each participant can see through their headset a visual representation of both the molecular simulation and all of the other participants (e.g, as in Fig 3). We designed the Isness-D experience to accommodate 4 – 5 participants wearing either the HTC Vive Pro or Valve Index HMD. Our cloud computing resources enabled us to mount our server on GPU shapes available at three different data centres: Frankfurt (Germany), London (UK), and Ashburn (Maryland, USA). Prior to entering Isness-D, participants at each node were fitted with the Open Mudra Gloves for Virtual Reality (OMG-VRs), etextile gloves that sense when a user pinches together their thumb and index finger, or thumb and middle finger to form a "mudra" position. (*13*) The OMG-VRs were designed to enable research scientists and students to 'grasp' and 'manipulate' molecular objects whose essence is purely energetic. They provide good positional tracking of the point at which a pinch takes place, require no calibration, and proved sufficiently robust to withstand use across the international network of Isness-D nodes in Fig 3C. The OMG-VRs played an important role during the Isness experience. By adopting a 'mudra pose' (bringing the tip of their thumb in contact with the tip of either their forefinger or middle finger), Isness-D participants could generate light and collectively sculpt the dynamics of the tangible molecular object. For the participants' bodies, the in-world renderers are designed so that the intensity of the light generated during moments of energetic coalescence depends on the overlap between energetic bodies. For example, two coalesced bodies generate more light than a single body, and four coalesced bodies generate more light than two.

### 3.3 Isness Nodes and Citizen Science Design

Each of the 14 different nodes distributed across the world (Fig 3C) was equipped with the equipment required to run Isness: 1) a room-scale play space of *2m x 3m*; 2) a VR-capable computer; 3) an HTC Vive Pro or Valve Index HMD; 4) a single pair of OMG-VRs; and 5) a reliable internet connection. Each node was managed by a 'host': an individual whom we trained over Zoom/Slack in how to set up the VR hardware and how to connect an Isness client into a cloud-mounted server instance. Hosts were given guidance on how to prepare participants prior to entering Isness-D (e.g. discussing what the experience would involve, creating a calming space, and offering participants an opportunity for quiet reflection), how to deal with participant discomfort, how to troubleshoot technology issues, and how to sanitize the VR equipment after participant use. In order to give hosts a sense of what their participants would experience during Isness-D, and also some indication of how the experience should feel, each of the hosts had the opportunity to undergo the full experience prior to engaging participants.

### 3.4 Participants & Data Gathering

Participants for Isness-D were recruited by each of the node hosts. After the node hosts collectively agreed a schedule of Isness-D time slots during which they were available to accommodate participants, this list of timeslots was placed into a web scheduler, with a link that was then circulated to participants. After receiving the web link, participants were able to identify the various session times available at their host node. Before registering for an Isness-D time slot, they first had to read the online study information sheet and complete a tick box consent form. We based our data collection on the structure of the psychedelic ceremony study, (*78*) which involves a 'pre-ceremony' and a 'post-ceremony' component and includes a combination of psychometrics commonly used in psychedelic research studies. The online pre-Isness-D questionnaire (using the survey platform Qualtrics) consisted of the ICS scale (*56*) and items relating to participants' openness to the upcoming experience. The post-Isness-D questionnaire included: 1) the ICS scale; (*56*) 2) the MEQ30 (*26, 27*) 3) the EDI; (*79*) and 4) the Communitas questionnaire (*61*). Additionally, participants were asked questions related to bodily effects and demographics. We added to the post questionnaire a *final comments* section where participants were able to leave written feedback. Since it was designed to evaluate YD experiences, not all aspects of these pre- and post-ceremony questionnaires were relevant to Isness-D. To ensure its relevance, we undertook a number of modifications. For example, we removed any questions asking about *suffering* and *surrendering*; we judged these to be less relevant to the Isness-D experience and moreover our preliminary analysis suggested that they increased anxiety among participants. The final versions of the pre- and post-Isness-D questionnaires used to evaluate Isness-D are given in the SM. Participants received an email link to a digital version of the pre-Isness-D questionaire approximately one hour prior to entering into Isness-D, and another link to the post-Isness-D questionaire after the conclusion of their session.

### 3.5 Isness-D Experience Design

Participants experienced Isness-D as a 3-phase journey, composed of a preparatory/induction phase, the Isness-D session, and an integration/reflection phase, adopting a similar model to that which we have utilized previously, and which is based on the model utilized in YD assisted psychotherapy. (*22*) An Isness-D session involved a facilitator along with 3 – 5 participants, each of whom simultaneously connected into the Isness server from one of the nodes shown in Fig 3. In some cases, additional observers were present within the Isness-D environment: by utilizing 'spectate' mode, they were invisible to the participants and to the facilitator. As technical assistants, observers played an important role in helping to stabilize the Isness-D experience. The 'spectate' function also helped us to train new facilitators, giving them the opportunity to observe first-hand without affecting the participant experience. All of the Isness-D nodes had available a play space of approximately 2m x 3m. To align participants within the shared virtual space, we adopted the following radial orientation strategy: we placed at the center of each node's space a vector that pointed toward one of the long (3m) boundaries, aligned each node's space so that their centers were coincident and the vectors were aligned, and then rotated each node's vector to make an angle of 360/n° with respect to its neighbors (where n is the number of participants). Fig 3 shows the resulting orientation for our most typical arrangement, with n = 4 participants.

Isness-D was constructed from a set of 'aesthetic hyperparameters', each of which controls some aspect of the participants' phenomenological experience and can be precisely varied using an available set of on-screen sliders. We defined a phenomenological 'state' as a given set of aesthetic hyperparameter values with some specified time duration. Each phase of Isness-D is comprised of a set of states, whose hyperparameter values are saved to a JSON file to ensure reproducibility. The Isness-D experience involved varying 69 different aesthetic hyperparameters, including for example: the color, distribution, density, and latency of the energetic bodies; the size of the heart center light; the rendering options for the energetic thread shown in Fig 3; options for setting interactive forces to achieve different effects; the state duration, and the global light levels. Our decisions on how to set the aesthetic hyperparameters were grounded in the various design considerations outlined in our previous work. (*25*) Each phase of Isness-D involved two audio components (played through the HMD earphones): 1) communication with the other participants in the space, which was achieved using an audio-only Zoom call using the mic built-in to the HMD; and 2) a pre-recorded narrative soundtrack, which we streamed as a separate audio channel using Twitch (www.twitch.tv). In what follows we describe the various phases of the Isness-D experience in further detail.

### 3.6 Pre-Experience Technical Checks

At the beginning of a scheduled Isness-D session, the node hosts would: 1) connect to the cloud mounted Isness server using an I.P. address provided by the session facilitator; 2) connect to an audio-only zoom call, using the earphones and mic in their HMD, and 3) connect to the Twitch feed. After the hosts confirmed that all of the relevant equipment was either plugged in or charged up, they would then enter into the Isness-D virtual environment with the facilitator prior to any participants entering and carry out: 1) an interaction latency check, to identify nodes whose latency or fluidity was significantly different from others as a result of a slow internet connection or network instability; 2) an OMG-VR check, verifying that the gloves generated light through the mudra pose, and were able to interact with the dynamical molecular object; 3) an audio check, where the facilitator ascertained whether any of the host nodes were suffering significant Twitch latency; 4) a spatial orientation check, where each node host confirmed that their respective radial orientation within the Isness-D virtual space minimized cable trip hazards. Hosts were asked to ensure that their participants were not in the same room while these checks were performed; however, this was not possible in many cases owing to space constraints.

We found Isness-D offered a fluid and stable experience so long as the internet connection at each node was stable. Persistent packet loss was the most substantial technological challenge to a smooth Isness-D experience. Technical instabilities were more common during the first week of our studies. Early on, we realized that the fluidity and reliability of an Isness-D experienced depended less on network speed and more on network stability (i.e., reliability low jitter, and low packet loss). We encouraged our nodes to evaluate their network stability using a simple online test, (*80*) and undertake optimization where possible – e.g., using a wifi router rather than tethering to a phone, moving closer to the wifi router, using an ethernet cable where possible, and encouraging other network users to avoid heavy usage during an Isness-D session. This combination of measures significantly reduced network-related glitches, creating a robust set of distributed nodes, and led to an increase in the number of sessions that we judged to be technologically seamless.

### 3.7 Phase 1: Preparation

Once the facilitator was satisfied that each node passed the technical checks outlined above, they invited each host to fit their participants with the HMDs and a pair of OMG-VRs, and ensure they were in a comfortable seated or kneeling position. When all the participants were ready, the facilitator initiated the Isness-D state sequence and Twitch audio stream. Over the next ~15 minutes (7 aesthetic states), the facilitator: 1) introduced themselves and greeted the participants; 2) asked each of the participants to introduce themselves, saying where they were located physically and how they were feeling; 3) instructed participants to alert their hosts if at any stage they suffered network problems or wished to leave; 4) thanked the participants for participating in our data collection efforts, and 5) asked that everybody respect the anonymity of the others. Next, the facilitator invited each of the participants to stand up and participate in some gentle movement exercises, to help them develop familiarity with moving in Isness-D and interacting with others. Participants were led in the following sequence: 1) acknowledging the others in the space by placing their hands in a prayer position and bowing to one another; 2) three rounds of gently raising and lowering their arms, with synchronized inhalation and exhalation; 3) rapport-building where each participant in turn was offered the opportunity to make up a motion which the others could then simultaneously mimic; and 4) encouraging each participant to gradually move toward the center of the virtual space ("a little closer… a little closer… a little closer…") until they all overlapped in order to have the phenomenological experience of coalescing their energetic body with the other energetic bodies (Fig 3A and 3B).

The facilitator then invited the participants to return to a comfortable seated or kneeling position, practice generating light using their OMG-VRs, and sculpt the dynamics of the molecular thread. Expanding the scale of the molecular thread, the facilitator encouraged all the participants to hold onto it, creating a circular shape (Fig 3A.i). The facilitator invited the participants to imagine Isness-D as a kind of meditation on the relationship between matter and energy, and how new senses of interconnectedness might emerge from imagining ourselves, others, and the world around us as luminous energetic essences rather than rigid material objects. The molecular thread was described to participants as a kind of tangible dynamical organism which – because its essence was purely energetic – could operate as kind of thread to connect our energetic essences across different physical locations and time zones. Whilst holding the molecule, the facilitator framed an intention-setting exercise, inviting each participant in turn to say a few words about something to which they would like to connect, inviting them to imagine the molecule as an energetic vessel to contain these intentions.

### 3.8  Phase 2: A loosely guided choreography

As the visual display of each participant faded to darkness, the facilitator went into 'spectate' mode, and another narrator took over. For the next 25 mins, participants were guided through a pre-recorded narrative soundtrack played through the HMD headphones. The narrative began with a short visualization, inviting participants to close their eyes and imagine their breath as radiant light concentrated at their heart center. After being instructed to open their eyes, the participants were guided through 16 states, each composed from a different combination of aesthetic hyperparameters. The narrative journey was an adaptation of the experience we previously developed for Isness-C, (*25*) but reworked in order to accommodate the unique affordances of the distributed environment. The experience took participants through a loose narrative of energetic emergence, dissipation, and fluctuation. It was designed to balance moments of individual introvertive exploration with collective extrovertive interaction. The narrative was accompanied by a soundtrack designed to broadly reflect the arc of the journey. Whilst in 'spectate' mode, the facilitator kept watch over the virtual space, in order to discern signs of potential technical difficulties, contact the relevant hosts in cases where there may be technical problems, and gently remind participants who became particularly excited that they should move gently, carefully and fluidly.

### 3.9  Phase 3: Integration & Discussion

Once the pre-recorded narrative finished and the participants were once again in a comfortable seated or kneeling position, the facilitator then re-emerged from spectate mode, making themselves visible to the Isness participants and gently greeting them. As during phase 1, the facilitator altered the scale of the energetic molecular organism and encouraged all the participants to hold onto it, forming an energetic circle which connected together the participants, as shown in Fig 3A.i. Over the next 10 mins, the facilitator encouraged the participants to share any thoughts, feedback, emotions, or impressions which arose during the experience. These group discussions were undertaken to enable maximum conversation amongst the participants, with the facilitator's role as that of an active listener. The format of these group discussions was not rigidly prescribed; instead, the emphasis was on enabling conversation amongst the group. In many groups, participants did not immediately wish to speak. In such cases, the facilitator asked a simple question to initiate the conversation – e.g., 'How do you feel?'; 'What remains?'; 'Were you aware of the others in the space with you?' Each of these discussions was recorded on Zoom, and subsequently transcribed as part of our data gathering efforts. After the discussion finished, the facilitator instructed participants in how to gently exit VR. Specifically, the participants were instructed to: 1) close their eyes; 2) gently remove their HMDs and place them on the ground next to them; 3) lie down; 4) let the light and sound in the ambient environment seep back into their conscious awareness; 5) open their eyes when they were ready; and 6) undertake the reflective writing on the paper and pencil provided from their host. This anonymized free writing was then photographed by hosts and sent to us for qualitative analysis. Those who were not ready to write in their post-Isness emotional state had the option to do so in their own time.

## 4 RESULTS

### 4.1 Sample Size, Demographics, and Emotional Response

From Aug-Sept 2020, we carried out 29 Isness sessions distributed across our network of 14 Isness nodes, with 109 total participants. 24 participants were not included in the analysis because their participation formed part of a technical training session or support role (i.e., as citizen science 'volunteers'). 9 participants were removed from the analysis because the facilitator or node host identified that their experience suffered significant technical difficulties – primarily glitches and dropouts as a result of network instability. This left 75 remaining participants whose Isness-D sessions could be analyzed. Of these, 54 answered the pre-Isness-D questionnaire, and 58 answered the post-Isness-D questionnaire. Of these 58 participants, 59% were male, 37% were female, 2% were other, and 2% declined to state. Ages ranged from 23 to 75 years old with a mean (SD) of 38 (10) years. Over half the participants (57%) had not been in VR at all over the last year. Four (7%) had used VR 20+ times in the last year. 12% had experienced the earlier, co-located version of Isness. (*25*) One reported a brief period of nausea (they did not comment on the strength or duration and finished their session as usual). Three reported body aches and three reported headaches. Six participants reported crying at some point during their Isness-D experience. Anecdotally, our study team obtained several other reports of people who experienced a brief period of tears. One participant reported trembling and two felt sweats or chills. None of the participants reported these emotional reactions to be negative or traumatic.

### 4.2 MEQ30

The MEQ30 (*27*) is a questionnaire that can distinguish dose dependent effects of YDs. It asks participants to rate the intensity with which they experienced 30 items on a 6-point scale [from "0 = none; not at all" to "5 = extreme (more than ever before in my life and stronger than 4)"], with three questions to capture ineffability **I**, fifteen mystical **M** [capturing unitive experiences, noetic quality, and sacredness], six positive mood **P**, and six transcendence of time/space **T**. Participant responses for each factor (**I**, **M**, **P**, **T**) are then averaged, and reported as a percentage of the maximum score. Fig 4 compares the Isness-D MEQ30 factor scores to 27 previous studies: 26 studies in the altered states database (*28*), where the MEQ30 has been used to analyse YD altered states, and our previous Isness-C study. (*25*) To analyze these results, we followed the approach Barsuglia et al. (*81*) used to analyze MEQ30 results obtained during MeO-DMT field tests. Specifically, we undertook comparative analysis of Isness-D to the previously published studies using independent sample t-tests with a significance level of 5% ($\alpha = 0.05$) calculated using Python (see SM for further details). Our previous work has shown that independent sample t-tests give results broadly aligned with more sophisticated statistical analyses. (*25*) Fig 4 and Table SM2 show the results of 27 different independent-sample t-tests, comparing the Isness MEQ30 results to each of the studies in Table SM2. To make Fig 4, we analyzed whether a study was statistically indistinguishable ($p > 0.05$) from Isness-D on each of its **I**, **M**, **P** or **T** factor scores. (*25*) Compared to Isness-D, Fig 4 shows:

- **3 studies which were more intense on all 4 factors**: 1) an MeO-DMT study (*81*); 2) a 30 mg psilocybin study [26]; and 3) a 2018 study by Griffiths et al. (*82*) where participants in a program offering high levels of support in carrying out meditation and spiritual (M/S) practice were given 20 – 30 mg psilocybin
- **6 studies which were more intense on 2 factors**: Five of these studies involved participants being administered doses between 20 – 42 mg of psilocybin. The sixth included our previous Isness-C study (*25*).
- **7 studies of equal intensity (indistinguishable on 3 or 4 factors)**: 3 studies were indistinguishable on all 4 factors (a 21 mg psilo study, (*83*) a 20 mg psilo study, (*84*) and a 200 μg LSD study (*85*)), and 4 were indistinguishable on three factors (psilocybin studies of 5 mg, (*14*) 10 mg, (*14*) 31.5 mg, (*83*) and a 200 μg LSD study (*85*)).
- **1 study which was less intense on 2 factors**, where participants were given dextromethorphan. (*84*)
- **10 studies which were less intense on 3 or 4 factors**: Amongst these, 6 were baseline studies, and 4 were drug-administration studies. The baseline studies included: 4 placebo studies; a 2016 study where Griffiths et al. gave participants sub-perceptual (1 – 3 mg) doses of psilocybin (*86*); and a 2018 study where Griffiths et al. (*82*) gave 1 mg psilocybin to participants in a program offering M/S practice support. The drug-administration studies investigated: MDMA (*85, 87*); methylphenidate (*85, 87*); ketamine (*88*); and psilocybin (*14, 84*).

Griffiths and co-workers identify an MEQ30 respondent as having had a 'complete MTE' when each of the **I**, **M**, **P**, **T** factor scores are ≥ 60% of the maximum. (*22*) In general, the fraction of participants reporting a complete MTE is proportional to the YD dose. Barrett and Griffiths (*89*) reported a meta-analysis of high dose (30 mg/70 kg) psilocybin studies in 119 healthy volunteers (*13, 14, 82*), and observed that **57%** of participants had 'complete' MTEs. Our analysis shows that **29%** of Isness-D participants had a complete MTE, compared to the previous value of **44%** for Isness-C.

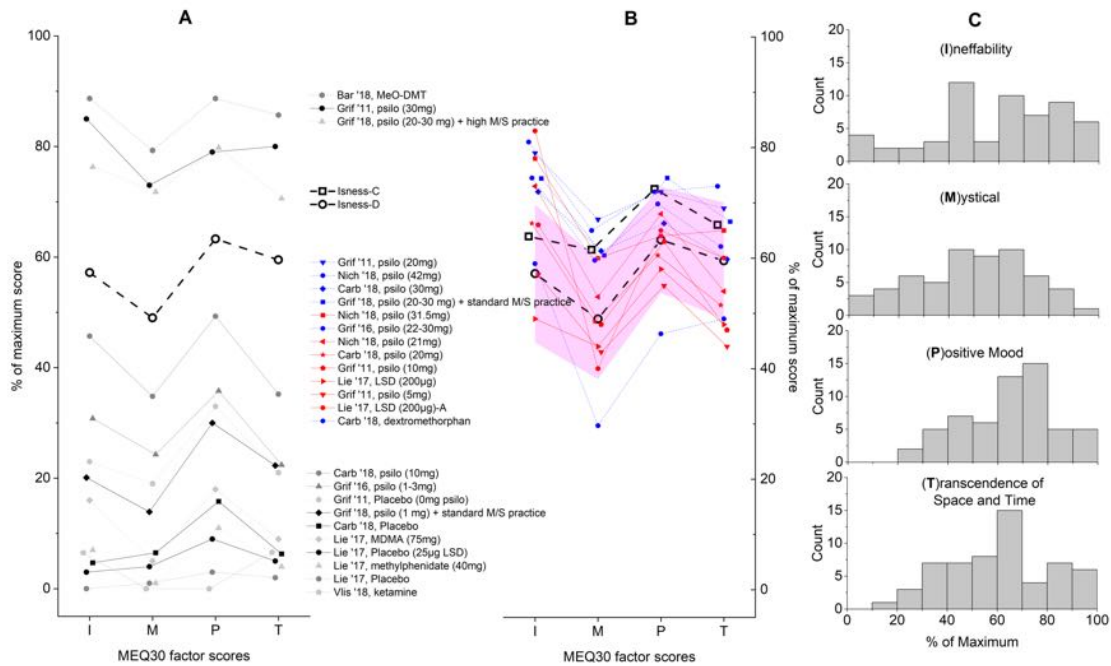

Figure 4: Comparison of the average (**I**, **M**, **P**, **T**) factor scores of Isness-D to Isness-C (*25*) and previously published research studies that have employed the MEQ30 (Bar '18 (*81*), Carb '18 (*84*), Griff '11 (*14*), Grif '16 (*86*), Grif '18 (*82*), Lie '17 (*85*), Nich '18 (*83*), Vlis '18 (*88*), and Isness-C (*25*)). (A) shows the average factor scores for the studies that were statistically distinguishable from Isness-D ($p < 0.05$). (B) shows studies that were statistically indistinguishable from Isness-D on at least 2 factors ($p > 0.05$), where the blue lines indicate a study with two factor scores indistinguishable from Isness-D, and the red lines indicate a study with three or four factor scores indistinguishable from Isness-D. The area highlighted in pink shows the approximate region within which a factor score was determined to be indistinguishable at a 5% confidence level. (C) shows the distribution of the four factor scores of the MEQ30 results of Isness-D.

### 4.3 Inclusion of Community in Self Scale

54 participants answered both the pre and post questionnaires, both of which included aforementioned ICS scale shown in Fig 5. The ICS asked which set of circles best describes their relationship with the other Isness participants from a series of images. The first image showed two circles with no overlap (a score of 0) and the last image showed two circles almost entirely overlapping (a score of 5). Fig 5 shows a significant shift in the distribution of ICS scores following Isness-D, with participants indicating higher degrees of connectedness. A Wilcoxon signed rank test showed that the difference between the pre and post values was statistically significant ($p < 1E-6$), with the mean ICS rating of $1.2 \pm 1.5$ (SD) pre-Isness-D increasing to $2.9 \pm 1.4$ (SD) post-Isness-D. The post-Isness-D ICS results are statistically indistinguishable from recent results published by Forstmann et al., (*90*) who investigated the effects of psychedelic substance use on positive mood and social connectedness of 450 participants (in naturalistic mass gathering settings) who used hallucinogens and/or euphorics in the last week. Given that the Forstmann et al. (*90*) scale ranged from 1 to 7, we mapped their results onto a 0 to 5 scale to enable comparison. Forstmann et al. found an average IOS of $2.8 \pm 1.3$ (SD), which is statistically indistinguishable from the post-Isness-D results according to independent sample t-tests ($p = 0.510$).

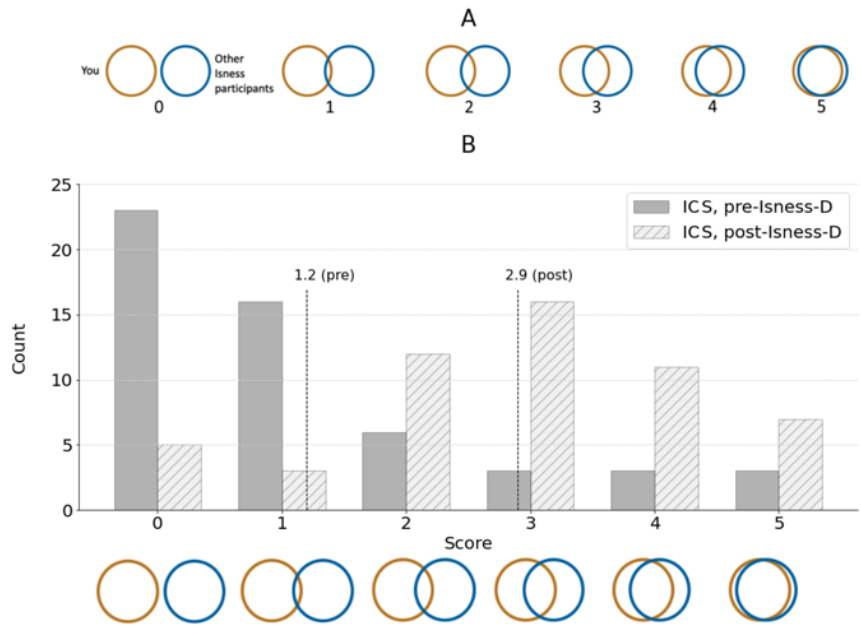

Figure 5: Panel (A) shows the ICS scale shown to participants in the pre-Isness-D and post-Isness-D questionnaire. Each picture has been labelled numerically from 0 to 5 for analysis, however, the pictures in the questionnaires were labelled alphabetically from (a) to (f). Panel (B) gives histograms of the ICS scores for all participants before and after Isness-D, with dashed lines indicating the respective averages.

### 4.4 Ego-Dissolution Inventory (EDI)

The EDI (*79*) self-report scale is typically employed in psychedelic studies to measure subjective experiences of ego dissolution. It consists of 16 items: 8 relate to ego dissolution (e.g., *I experienced a dissolution of my "self" or ego", I felt a sense of union with others, etc.*) and 8 relate to ego inflation (e.g., *I felt more important or special than others*, *I felt especially sure-of-myself, etc.*). Following Isness-D, participants were asked to rate the extent to which each statement applied to their Isness-D experience on a scale of 0 (*no, not more than usually*) to 100 (*yes, I experienced this completely)*. Cronbach's alpha was calculated, and the results showed high internal consistency for both ego dissolution ($\alpha = 0.86$) and inflation ($\alpha = 0.87$). Fig 6 shows that the ego inflation distributions are shifted to lower values compared to the ego dissolution distributions, with a mean ego dissolution across the cohort of $40 \pm 20$ (SD), and mean ego inflation of $17 \pm 15$ (SD). The results in Fig 6 are consistent with subjective participant reports following LSD-equivalent doses of ~100 μg, based on inspection of values published by Nour et al. (*79*) Direct comparison of the complete Isness-D EDI scores to previous YD studies is complicated owing to the fact that several previous studies use only a single EDI item to measure ego dissolution, usually *I experienced a dissolution of my "self" or ego*, or similar. For Isness-D, the mean for this specific question was $43 \pm 29$ (SD), which is higher ($p < 0.001$) than the subjective responses following 75 microgram doses of LSD published by Tagliazucchi et al. (*57*) and lower ($p < 1E-14$) than responses following 100 micrograms of LSD, published by Holze et al. ($83 \pm 10.2$). (*58*) In the same study, Holze et al. reported the mean ego dissolution across participants following a 125 mg dose of MDMA to be $44 \pm 7.9$ (SD), which was found to be indistinguishable ($p = 0.702$) from Isness-D. Madsen et al. (*59*) used the EDI to measure ego dissolution for participants following a range of doses of psilocybin (0 – 30 mg). Compared to Isness-D, Madsen et al. found similar or lower mean ego dissolution values following doses up to and including 18 mg of psilocybin. Increasing the psilocybin dose further led to higher mean scores for all participants compared to Isness-D. Mason et al. (*60*) measured the subjective ego dissolution of participants following 0.17 mg of psilocybin per kg of body weight (avg ~12 mg dose per person) using the EDI. The mean across the participants was 30.7 ± 4.6 (SD), which is lower (p = 0.003) than Isness-D.

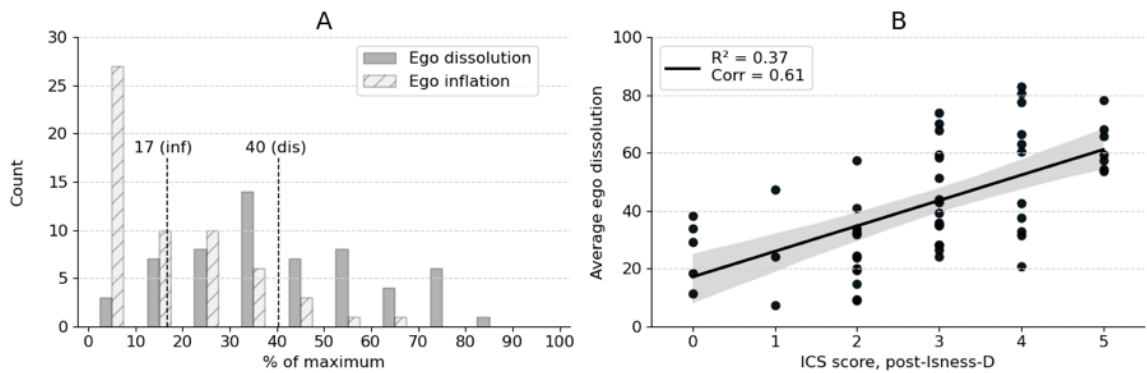

Figure 6: (A) distributions of the average scores for the ego inflation and ego dissolution items of the EDI across the Isness participants. The respective averages are indicated by the dashed lines. (B) scatter plot of the average ego dissolution against the post-Isness-D ICS scores for each participant, with a line of best fit calculated using a linear least-squares regression. The legend gives the Pearson correlation coefficient ('corr') and the $R^2$ value.

To evaluate the relationship between a participant's sense of ego dissolution and their sense of connectedness, Fig 6B shows a correlation plot of each participant's average post-Isness-D ego dissolution against the corresponding post-Isness-D ICS score. While there is some scatter in the data, a linear least-squares regression indicated a positive correlation ($R^2 = 0.37$) between the average post-Isness-D EDI score and the ICS score. The Pearson correlation coefficient (0.61) similarly indicated a positive correlation. These results are distinct from the weaker correlations observed between the post-Isness-D ego dissolution scores and the pre-Isness-D ICS scores ($R^2 = 0.17$ and a Pearson correlation coefficient of 0.41). The change in correlation suggests that the sense of social connectedness which Isness-D elicits for participants correlates with a sense of ego attenuation.

### 4.5 Communitas

Following Kettner et al., (*61*) we used the scale in Fig 7 to assess participants' Isness-D experiences. The first 8 items offer a subjective assessment of communitas, item 9 assesses participant-to-participant connection, and item 10 assesses participant-to-facilitator connection. The scale ranges from 1 (*strongly disagree*) to 7 (*strongly agree*). Calculation of Cronbach's coefficient showed high internal consistency within the 10 items ($\alpha = 0.84$). A one-sample t-test against the midpoint of the scale showed that the overall mean of the 10 items across the participants was higher than the midpoint of the scale value (p < 0.001), indicating that a sense of togetherness was felt by participants during their shared Isness-D sessions. The item *the ceremony really allowed me to get to know the other participants* was the only one to receive a lower mean score than the midpoint scale value, perhaps because the Isness-D experience involved little verbal communication between participants. The rest of the items all received ratings higher than scale midpoint value, with the item *during the ceremony, I felt that social status became irrelevant* receiving the highest mean rating. In recently published studies where Kettner et al. explored *collective* psychedelic use during YD ceremonies in naturalistic settings (N = 886), they calculated the mean reported total communitas as the sum across the first 8 items in Fig 7. They obtained values of 39.58 (SD 11.23), corresponding to ~71% of the maximum score of 56. Isness-D produced mean reported communitas values that are statistically distinguishable (p = 0.002), with an average total communitas of 44.14 (SD 6.87), which corresponds to ~78% of the maximum score.

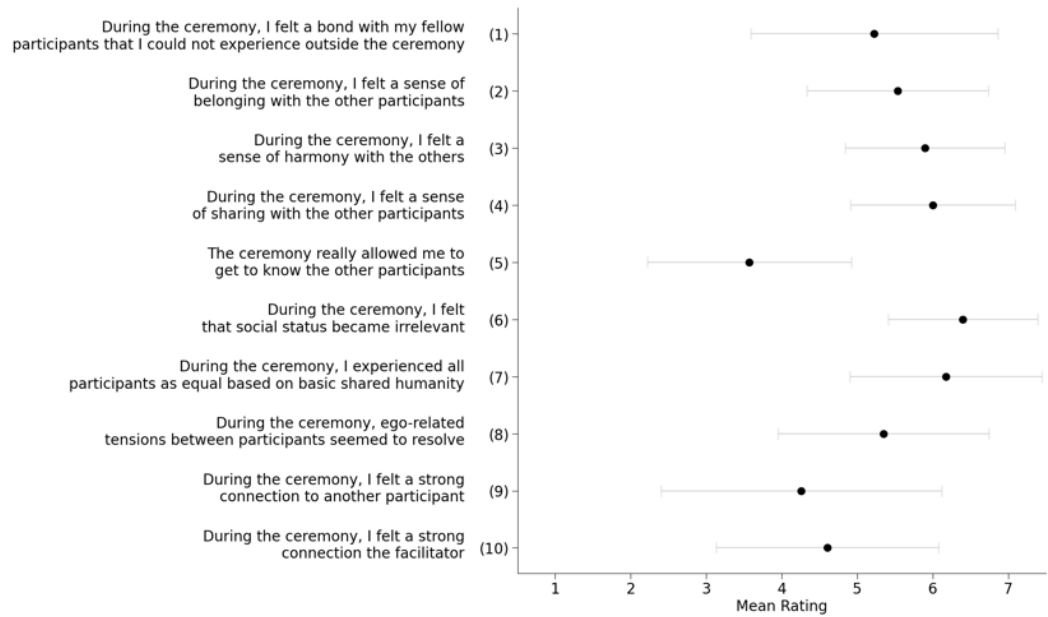

Figure 7: Mean ratings for the 8 items in the Communitas questionnaire (items 1–8), and two additional items relating to connection with another participant and with the facilitator (items 9 & 10), with ±1 SD error bars.

### 4.6 Qualitative Analysis

Our qualitative analysis is based on data from three sources: 1) semi-structured group discussion in VR during phase 3 of Isness-D; 2) optional free writing upon exiting VR; and 3) a post-Isness-D questionnaire. All the group discussions were recorded (~3.5 hours total) and then transcribed (~14,000 words). We received 36 free writing responses in a variety of forms: prose (27 legible and 1 illegible), poetry (4 responses), and drawings (4 responses). Only the prose (~3000 words) was included in the thematic analysis. An inductive thematic analysis (*91*) was used to categorize the qualitative data into themes. Any statements that involved multiple themes were assigned to the most relevant. The identified themes and the number of times each theme was observed are as follows: Connectedness (75); Positive Emotions (65); Embodied awareness (40); Ego Dissolution (31); Supportive Setting (28); Sense of Play (24); Transcendence of Space/Time (13); Noetic Quality (11); Comparison to other Altered States (10); Sense of Beauty (9); and Reflection on Mortality (8). Table SM6 contains indicative quotations for each theme; the SM includes the entire thematic analysis classification. Several of the participant statements (*written in italics*) are woven into the discussion that follows.

## 5 DISCUSSION

Our results overwhelmingly suggest that the phenomenological intensity of STEs which arose for participants during Isness-D is comparable to YDs, in both naturalistic and laboratory settings. For example, Isness-D participants' average ICS score of 2.9 ± 1.4 (SD) is statistically indistinguishable from recent results published by Forstmann et al., (*90*) who found values of 2.8 ± 1.3 (SD) in a large-scale naturalistic study investigating YD effects for more than 450 participants. Similarly, the extent of communitas reported by Isness-D participants (44.14 ± 6.87, corresponding to ~78% of the max score) is statistically higher than the value of 39.58 ± 11.23 (~71% of the max score) reported in Kettner et al.'s recent N = 886 study investigating *collective* psychedelic use during YD ceremonies in naturalistic settings. Our EDI analysis indicates that the Isness-D results are comparable to YD drug does of ~18 mg psilocybin, 75 – 100 μg LSD, or 125 mg MDMA. These results are broadly consistent with the results of the MEQ30 analysis, showing 3 previous YD studies which were indistinguishable from Isness-D on all 4 factors: psilocybin studies of 20 mg (*84*) and 21 mg, (*83*) and an LSD study of 200 μg (*85*). The qualitative analysis indicates some phenomenological similarity between Isness-D and psychedelics, with participants for example observing how it was "*similar to experiences that I have had as somatic visions through medicine plants. The interconnective nature of energy/intention and the 'strings' that appear to interconnect us with all living matter [is] also related to childhood dreams I had prior to any 'psychedelic experience'* ". Others stated how Isness-D left them with a "*sense of interconnectedness… only previously noticed with the help of psychedelics in the right setting*". Some commented on the emotional impact of Isness-D, "*I was amazed [by] how moved I was. I think it was the juxtaposition of the frustration and then the beholding beauty. I think the music really helped. It felt very beautiful to sit back and witness. I don't normally get moved quite so much.*" Similar to YD experiences, some participants attributed a spiritual significance to Isness-D, "*That's definitely a spiritual experience of some sort. It's like, tangible*". Because Isness-D grounds emergent spiritual concepts in rigorous physics insights like the fundamental relationship between matter and energy, it offers a case study for how careful human computer interaction (HCI) design can be used to cultivate spiritual sensitivities which avoid the "woo-woo" pseudo-scientific associations which can arise from YDs, e.g., which have recently been discussed by Carhartt-Harris and Friston. (*15*)

Connectedness, which has been previously been highlighted as a key aspect of the YD experience, (*92*) emerged as the strongest qualitative theme for Isness-D participants, "*I felt connected with myself but also with everyone else here… I think 'connected' is the word for me for the end of this session.*" Others commented on how Isness-D offered "*A completely other way of connecting that I'm not familiar with, [where] all the usual stuff disappears.*" Comments like these are aligned with the observed change in ICS post-Isness versus pre-Isness. The matter-energy narrative of Isness-D and its aesthetics of luminosity – which blurred the boundaries of individual identity – seems to have been important for enabling participants to reimagine a sense of connectedness: "*Connecting light and emotions as energy was special… light as connection; light transferring between matter; light creating memory. Experiencing myself and the other people in the group as light energy was joyful. It… allowed me to think about other spaces connecting in the world.*" While the experience of energetic coalescence was only one aspect of the Isness-D experience, it seemed to be particularly poignant and intimate for many people, with participants "*struck with how quickly the abstract lights grew to hold tangible meaning… as other people. It was especially poignant [during moments of coalescence] when we moved toward the centre and felt as if we might collide or enter each other's personal space.*" For many participants, coalescence facilitated a particularly strong sense of connectedness: "*we could get closer than [in real-life] which felt more intimate, and connecting—nearly as much so as with a partner, child or pet—even though we were in different places.*" Others observed how moments of coalescence produced somatic sensations: "*I can just literally walk into people and it's quite sensuous*", and also "*I could feel subtle changes in my hands as if something was passing by, something physical.*" Another commented how, during moments of coalescence "*I got quite emotional… I got this surge of emotion where I don't know if I wanted to gasp or cry or what it was, but I was kind of shocked in awe.*" Not a single participant was uncomfortable when others coalesced with their energetic essence; however, some worried they were invading others 'personal space': "*You can sense an imaginary presence around the glowing light, and you give it space even though there is no barrier there, other than imagined.*" Several commented on the sense of intimacy that accompanied coalescence "*It felt really intimate. And I tried to connect to whichever light being it was and… I tried to join mudras, and I think I sat [as an energetic essence] inside a few of you. I hope that was OK*" Many people reported a pleasant sense of spaciousness at the moment of coalescence, recognizing in that instant that conventional material boundaries did not apply, "*I found it really strange at the start, when you were telling us to go closer and closer and then that weird* *boundary of where personal space is and all of a sudden it was like 'oh, actually this is kind of nice!'* " Several participants commented on how the intimacy of the coalescence experience was balanced with a sense of innocence and purity, enabling "*the sweetest tenderness or pure, childlike love....*"

Earlier in this article, we highlighted two phenomenological ingredients known to be important in STEs: 1) diminished self-salience arising from dissolving self-boundaries; and 2) a sense of unity with something beyond the self. Given the tight relational co-dependence of these ingredients (where one is required to characterize the other and vice-versa), we suggested that they are *best realized in a multi-body context*, given that the softness, fluidity, or porosity of a body's boundaries become most apparent by observing its relational interactions with another body. One participant succinctly articulated how they experienced this relational co-dependence during Isness-D: "*you lose yourself in the connection to other people*". Fig 6B and Fig SM1 show that the correlation between participants' sense of connectedness and their diminished sense of egoic identity was significantly stronger following Isness-D than it was beforehand: the diminished sense of ego entailed in "losing oneself" arises through the unique sort of "connection to other people" that Isness-D enables. The relational co-dependence between connectedness and diminished egoic identity is also seen in quotations like, '*Identity didn't matter anymore; it was about experiencing things together. That was wonderful.*' As energetic essences, the implicit, explicit, conscious, and unconscious judgments that permeate social and relational interactions were diminished: '*I found [the anonymity] quite powerful because we were all completely equal in the space. Any of the pre-judgements that come from how people look, sound and that sort of thing just aren't there.*' Stripping away the various ego-games that characterize interpersonal interactions led to a sense of joy, purity, love, and beauty: '*I felt [when touching one another's heart centres] the sweetest tenderness or pure, childlike love… stripped back, without any of the assumed layerings that we place upon reality and relationships… just to the absolute core, it was truly beautiful.*' These comments support the exceptionally high scores observed for item 6 in the Communitas scale in Fig 7 (*During the ceremony, I felt that social status became irrelevant*). Yaden et al (*5*) note that the experience of self-loss is sometimes linked to pathologies (e.g., going back to the work of Freud); however, they argue that it is more often associated with positive outcomes. Our results are broadly aligned with this conclusion: the experience of self-loss for Isness-D participants was overwhelmingly positive.

Participants reported positive emotions and an overwhelming sense of calm and relaxation at the end of their Isness-D experience, "*I feel remarkably happy. There's something very happiness-inducing about this whole practice.*" Some reported lower levels of anxiety and stress, observing that Isness-D helped them to "*dissociate from something within you…I don't know if it… was my identity or my anxieties that I put on the side...*" Another similarly commented, "*I arrived feeling pretty anxious and disconnected from others, but after that experience I feel much more calm and hopeful. Hugely due to the focus of how energy doesn't truly disappear. I guess the fear of dying (the existential threat) is always on our minds, but that experience quieted the fear.*" For some, Isness-D encouraged reflection on mortality, and their own energetic transience amidst a larger unfolding energetic process: "*It's clear that the energy isn't just gone [in death], it goes somewhere. It's beautiful to think that it's all out there somewhere and it's still circling you constantly*" and also "*Seeing all the lights together reminded me of a dream I had years ago where the universe was ending and everybody was dissolving into balls of light and merging into one. So, I was like 'oh cool! The universe is ending. I'm cool with that.'* " Some participants articulated how Isness-D offered a form of connectedness that contrasts COVID-related isolation: "*[the sense of connection] is an experience I haven't felt in a while because of COVID and being in front of screens all of the time"*, and also "*I feel like we've had a hug, and I haven't had many of those recently… a really nice thing to have*". Another commented on how it helped them understand the difference between "*stillness and stagnation…. This [pandemic] can make you feel very stagnant, cause you're in the same place and you're not still or rooted to anything.*" For many participants, positive emotions seemed to be associated with the sense of playfulness that arose from relaxed egoic identity: "*It was…a space to be free, you could do whatever you want and not feel judged"* and also "*There certainly was a sense of playfulness in this, which was really nice. As adults that's not always something that we pursue.*"

For the MEQ30, the Isness-D and Isness-C results for the **I** and **T** factors were statistically indistinguishable. For the **M** and **P** factors, the Isness-D scores were lower than Isness-C. The lower scores make sense for several reasons. First, the distributed 'citizen science' approach meant that we had significantly less control over participant's set and setting prior to entering VR and after leaving VR. Whereas in our previous study we were able to offer all participants the same context, the same preparatory environment, similar psychological priming, and a more consistent experience of the technology, this was not possible using the distributed citizen science approach, where the variability was more significant. For example, we had little ability to influence participant expectations or control how the node hosts described Isness-D to participants during the recruitment phase. Especially in the early stages, there was variability in host preparation: for example, there were some cases where participants from one node would be ready and waiting in VR for the beginning of Isness-D phase 1, whilst other node hosts were troubleshooting the technology (e.g., HMD fit, focus, OMG-VR fit, etc.). Second, whereas our previously published results were all obtained using a local area network where we could closely monitor latency and quickly solve any technical problems, the same was not true for the work described herein. During the early stages of our citizen science study, a number of participants had intermittent disruptions to their plausibility illusion, (*40, 41*) as a result of unoptimized network instability issues. Third, whereas the previous participant cohort were drawn from attendees at a psychedelics & consciousness conference, the cohort for Isness-D was drawn from a broader distribution of international participants. Finally, Isness-D was effectively an adaptation of an experience that we originally designed to be co-located. Whilst we made several changes to accommodate the affordances of a multi-person distributed experience, our limited resources during the lockdown period made it difficult for us to undertake more extensive changes.

Isness-D differs from conventional YD psychotherapy in various ways. For example, the phase 1 preparation lasted ~15 mins, far less than the preparation for studies carried out for YD psychotherapy, which typically include a total of 4 – 8 individual sessions (both before and after the YD session). The three phases of Isness-D last a total of ~70 minutes, shorter than psilocybin and LSD experiences, which can often last from 6 – 14 hours. Finally, Isness-D was designed as a group experience, whereas most YD and VR studies are individual experiences. As discussed above, the distributed multi-body group aspect of Isness-D is clearly important for weakening ego boundaries, and fostering a sense of connectedness. The opening and closing interactions between the group and the facilitator (e.g., to develop the energy-matter narrative and encourage moments of energetic coalescence) were important, *'really help[ing] to hold the experience and make me feel more comfortable with everyone.'* Compared to previous YD studies, the N = 58 Isness-D sample size was reasonably sized; however, this study had a number of limitations. For example, while we were able to demonstrate significant changes post-Isness compared to pre-Isness, we did not carry out a control experiment, and therefore some of our analyses are primarily comparative. Whilst the 'citizen science' participant sample in this work clearly represents an improvement on our previous work, further work will be required to make definitive statements about the extent to which sample selection bias may have influenced our results. Our ability to compare the MEQ30 from this work with previous studies depends on the assumption that the baseline responses of our participant sample are not anomalously high or low, and within the range spanned by 6 previously published baseline studies. Fig 4 shows that these baseline studies have a broad MEQ30 score distribution. Our comparative statistical analyses (Fig 4 and Table SM2) show that the Isness-D results are more intense (p < 1E-6) than all of them. Nevertheless, we believe that the results described herein provide a degree of confidence in the results which we obtained in our previous study. As discussed above, the fact that the MEQ30 results for Isness-D are slightly lower than the results obtained for the co-located Isness-C experience is entirely aligned with our expectations, given the very different circumstances in which each study was conducted. At this stage, it is unclear how exactly to define a "placebo" for an experience like Isness-D, but this is an issue that would be interesting to investigate in future work. The design of Isness-D as a group experience suggests that the individual data may be correlated. In future studies we wish to investigate the correlation of the results obtained for participants within specific groups, and compare *intra*-group results to *inter*-group results.

# 6 CONCLUSIONS

The aim of the work described herein was to determine whether the unique affordances of multi-person distributed VR can be used to reliably elicit intense STEs. Within this paper, we have described Isness-D, an experience we have developed to blur conventional self-other boundaries using the unique affordances of distributed multi-person VR. Built on a matter-energy narrative, Isness-D enables groups of participants distributed across the world to co-habit a shared virtual space and collectively experience their bodies as luminous energetic essences with soft spatial boundaries. It encourages participants to imagine themselves, others, and the world around them as unfolding interconnected processes which are energetic (rather than fixed material entities). This fluid energetic representation enables participants to undergo moments of 'energetic coalescence', a new class of embodied intersubjective phenomenological experience whereby participants can have an embodied experience of including multiple others within their self-representation. To evaluate Isness-D, we adopted a 'citizen science' approach, coordinating a network of nodes distributed around the world to run multiple Isness-D sessions. This strategy enabled us to carry out this research amidst COVID related social distancing constraints. As a distributed approach, it was difficult to prime Isness-D participants in a controlled way. Nevertheless, analysis of Isness-D participant scores on four different self-report scales commonly used to assess YD experiences overwhelmingly suggest that the phenomenological intensity of STEs which arose for participants during Isness-D is comparable to previously published YD studies, in both naturalistic and laboratory settings. Isness-D enables participants to dissolve their sense of self in the connection to others, relaxing attachment to egoic identity and facilitating a strong sense of connectedness. For many participants, Isness-D offered a sense of intimacy, innocence, playfulness, and purity, eliciting a state of calm spaciousness. To the best of our knowledge, this work represents the first attempt to analyze a distributed multi-person VR experience using measurement scales which are typically applied to YD experiences. These results demonstrate that distributed VR can be used to elicit intersubjective STEs which simultaneously attenuate egoic identity and facilitate a sense of connectedness. This study reaffirms the ideas in our previous work, (*25*) where we speculated that it is possible to design phenomenological experiences using *numadelic* technologies like multi-person VR to create the conditions for STEs from which participants derive insight and meaning. Distributed intersubjective VR experiences like Isness-D may have a role to play in easing unprecedented feelings of loneliness, isolation, and fear that have arisen with COVID restrictions. In future work, we hope to explore in futher detail the mechanisms responsible for the results obtained during Isness-D, and also carry out detailed follow-ups with participants in order to understand its impacts over the longer term.

**ACKNOWLEDGMENTS**

*Author Contributions:* DRG led the project, conceived the idea, organized and supervised the team, wrote the original draft, co-designed the Isness-D experience, guided the software design, and provided funding. RRW ran an Isness node, co-designed the experience, analyzed the data, and helped write the paper. MDW worked on the software and helped make figures. OMM ran an Isness node, helped carry out participant studies, assisted in data collection, and commented on the paper. RF made the mudra gloves, ran an Isness node, helped carry out the participant studies, and helped make the figures. JEP and MC co-designed the Isness-D experience, helped carry out the participant studies, ran two Isness nodes, and organized the Isness node network. *Funding:* This project was supported by the Leverhulme Trust (Philip Leverhulme Prize), Royal Society (URF/R/180033) and EPSRC (EP/P021123/1); RRW is funded by the EPSRC TMCS CDT (EP/L015722/1); MDW by the Royal Society (RGF/EA/181075); JEP, MC, & RF by the ArtSci International Foundation. *Thanks:* We would like to thank the network of Isness participants, node hosts, and citizen science volunteers around the world who donated their time to this project, and without whom we could not have accomplished this work. We specifically acknowledge the following people: Gregory Roufa and Joseph Hardy ran Isness nodes and helped to organize and expand the network of Isness nodes. Matt Rojo, Justin Wall, Harrison Blum, Steph Suddel, Owen Williams, Vadmin Zhuk, William Zhuk, and Jeremy Lam, all helped to run Isness nodes. Kara de los Reyes and Manda Baynes helped with various aspects of the organization along the way, and Oussama Metatla made valuable comments on the paper structure. Isolde Freeth-Hale performed, recorded, and co-designed the audio component of Isness-D. The open source VR software used to construct Isness has been developed and maintained by various members and collaborators of the Intangible Realities Laboratory (IRL) over the years. The ability to run cloud-mounted instances of the Isness-D server was made possible through work by Dr. Jonathan Barnoud and Mark Pope, along with a generous grant of cloud computing credits from Oracle through their University Partnership Cloud Program.

Table SM1: Showing the mean MEQ30 (**I**, **M**, **P**, **T**) factor scores (as a percentage of the maximum score) for each participant.

| Participant number | Ineffability | Mystical | Positive Mood | Transcendence of Space & Time |
|---|---|---|---|---|
| 1 | 60 | 53 | 57 | 60 |
| 2 | 73 | 67 | 83 | 87 |
| 3 | 60 | 77 | 83 | 83 |
| 4 | 40 | 45 | 63 | 57 |
| 5 | 53 | 65 | 87 | 80 |
| 6 | 100 | 61 | 90 | 67 |
| 7 | 53 | 35 | 60 | 40 |
| 8 | 80 | 35 | 77 | 60 |
| 9 | 33 | 55 | 37 | 60 |
| 10 | 40 | 53 | 57 | 67 |
| 11 | 100 | 51 | 87 | 60 |
| 12 | 0 | 45 | 53 | 37 |
| 13 | 80 | 55 | 63 | 63 |
| 14 | 93 | 79 | 87 | 87 |
| 15 | 60 | 63 | 60 | 37 |
| 16 | 67 | 48 | 67 | 67 |
| 17 | 100 | 97 | 100 | 100 |
| 18 | 40 | 52 | 73 | 63 |
| 19 | 80 | 75 | 77 | 63 |
| 20 | 73 | 75 | 73 | 83 |
| 21 | 20 | 27 | 40 | 30 |
| 22 | 40 | 37 | 60 | 57 |
| 23 | 7 | 17 | 20 | 23 |
| 24 | 40 | 27 | 40 | 43 |
| 25 | 53 | 44 | 70 | 37 |
| 26 | 60 | 60 | 67 | 53 |
| 27 | 80 | 71 | 77 | 90 |
| 28 | 7 | 13 | 40 | 20 |
| 29 | 40 | 48 | 63 | 83 |
| 30 | 73 | 32 | 70 | 33 |
| 31 | 47 | 11 | 43 | 40 |
| 32 | 73 | 64 | 77 | 77 |
| 33 | 100 | 83 | 77 | 93 |
| 34 | 47 | 31 | 43 | 53 |
| 35 | 7 | 21 | 40 | 40 |
| 36 | 73 | 60 | 70 | 70 |
| 37 | 73 | 48 | 67 | 67 |
| 38 | 40 | 56 | 70 | 47 |
| 39 | 60 | 60 | 60 | 60 |
| 40 | 67 | 41 | 57 | 57 |
| 41 | 27 | 71 | 77 | 53 |
| 42 | 47 | 13 | 40 | 20 |
| 43 | 87 | 87 | 97 | 100 |
| 44 | 100 | 83 | 90 | 100 |
| 45 | 67 | 52 | 63 | 57 |
| 46 | 13 | 3 | 27 | 30 |
| 47 | 33 | 8 | 37 | 13 |
| 48 | 60 | 60 | 67 | 63 |
| 49 | 60 | 53 | 70 | 63 |
| 50 | 80 | 49 | 73 | 83 |
| 51 | 40 | 25 | 37 | 50 |
| 52 | 87 | 48 | 67 | 47 |
| 53 | 40 | 29 | 37 | 43 |
| 54 | 73 | 43 | 50 | 70 |
| 55 | 33 | 27 | 37 | 67 |
| 56 | 87 | 63 | 70 | 73 |
| 57 | 13 | 5 | 57 | 33 |
| 58 | 80 | 87 | 97 | 90 |

**Table SM2**: The *p* values from independent sample t-tests comparing the mean MEQ30 (**I**, **M**, **P**, **T**) factor scores from Isness-D to Isness-C and to previously published YD studies, calculated by the function "scipy.stats.ttest_ind_from_stats()" (Python 3.7 and SciPy 1.6.2). Studies highlighted in blue were statistically distinguishable for 1 or fewer **I**, **M**, **P** or **T** factor scores and more intense (i.e., higher mean MEQ30 score) than Isness-D. Studies highlighted in light grey are those for which two or more **I**, **M**, **P** or **T** factor scores were statistically indistinguishable from Isness-D. All values that satisfy $p > 0.05$ are in bold and indicate that the corresponding study results were statistically indistinguishable from Isness-D at a significance level of 5%.

| | | Ineffability | | | Mystical | | | Positive mood | | | Transcendence of Space & Time | | |
|---|---|---|---|---|---|---|---|---|---|---|---|---|---|
| | n | mean | std dev | *p* value | mean | std dev | *p* value | mean | std dev | p value | mean | std dev | *p* value |
| **Isness-D** | 58 | 57.2 | 26.6 | n/a | 49 | 22.4 | n/a | 63.3 | 18.7 | n/a | 59.5 | 21.6 | n/a |
| Isness-C | 57 | 63.9 | 20.4 | **0.13287** | 61.5 | 16.5 | 0.00092 | 72.5 | 11.4 | 0.00191 | 66 | 13.4 | **0.05547** |
| Bar '18, MeO-DMT | 20 | 88.7 | 12.6 | 0.00000 | 79.3 | 18.4 | 0.00000 | 88.7 | 11.7 | 0.00000 | 85.7 | 13.8 | 0.00000 |
| Grif '11, psilo (30mg) | 18 | 85 | 25 | 0.00019 | 73 | 25 | 0.00024 | 79 | 21 | 0.00344 | 80 | 25 | 0.00113 |
| Grif '18, psilo (20-30 mg) + high M/S practice | 25 | 76.3 | 19.5 | 0.00179 | 71.8 | 18 | 0.00002 | 79.8 | 19 | 0.00043 | 70.6 | 18.5 | 0.02796 |
| Grif '11, psilo (20mg) | 18 | 79 | 25 | 0.00291 | 67 | 21 | 0.00346 | 72 | 25 | **0.11682** | 69 | 30 | **0.14317** |
| Nich '18, psilo (42mg) | 12 | 81 | 26 | 0.00609 | 65 | 35 | 0.04646 | 72 | 28 | **0.18513** | 73 | 31 | **0.07303** |
| Carb '18, psilo (30mg) | 20 | 72 | 18.8 | 0.02456 | 61.3 | 21 | 0.03471 | 66.3 | 20 | **0.54511** | 59.8 | 17 | **0.95525** |
| Grif '18, psilo (20-30 mg) + standard M/S practice | 25 | 74.4 | 31.5 | 0.01250 | 60.5 | 36.5 | **0.08258** | 74.5 | 21.5 | 0.01908 | 66.6 | 28 | **0.21369** |
| Nich '18, psilo (31.5mg) | 11 | 78 | 26 | 0.01988 | 60 | 35 | **0.18009** | 64 | 29 | **0.91788** | 65 | 31 | **0.47436** |
| Grif '16, psilo (22-30mg) | 50 | 74.5 | 26 | 0.00093 | 59.6 | 29.8 | 0.03758 | 69.8 | 27 | **0.14453** | 62.1 | 23.9 | **0.55396** |
| Nich '18, psilo (21mg) | 10 | 73 | 27.2 | **0.08842** | 53 | 37 | **0.64050** | 68 | 28 | **0.49964** | 54 | 33 | **0.49635** |
| Carb '18, psilo (20mg) | 20 | 66.3 | 23.3 | **0.17803** | 48.5 | 28 | **0.93597** | 60.5 | 24 | **0.59372** | 51.5 | 21 | **0.15449** |
| Grif '11, psilo (10mg) | 18 | 66 | 25 | **0.21782** | 48 | 25 | **0.87255** | 63 | 25 | **0.95651** | 47 | 25 | 0.04235 |
| Lie '17, LSD (200µg) | 11 | 49 | 26.5 | **0.35166** | 44 | 26.5 | **0.51192** | 58 | 26.5 | **0.42453** | 48 | 23.2 | **0.11415** |
| Grif '11, psilo (5mg) | 18 | 57 | 30 | **0.97850** | 43 | 21 | **0.31728** | 55 | 30 | **0.16278** | 44 | 30 | 0.01823 |
| Lie '17, LSD (200µg)-A | 16 | 83 | 12 | 0.00034 | 40 | 24 | **0.16539** | 65 | 16 | **0.74137** | 60 | 20 | **0.93391** |
| Carb '18, psilo (10mg) | 20 | 45.7 | 19.4 | **0.08002** | 34.8 | 19.8 | 0.01404 | 49.3 | 18 | 0.00468 | 35.2 | 16.2 | 0.00002 |
| Carb '18, dextromethorphan | 20 | 59 | 18 | **0.77973** | 29.7 | 21 | 0.00117 | 46.3 | 22.1 | 0.00128 | 49 | 21.2 | **0.06349** |
| Grif '16, psilo (1-3mg) | 50 | 30.8 | 31.7 | 0.00001 | 24.3 | 27.1 | 0.00000 | 35.8 | 28.3 | 0.00000 | 22.4 | 20.5 | 0.00000 |
| Grif '11, Placebo (0mg psilo) | 18 | 23 | 21 | 0.00000 | 19 | 21 | 0.00000 | 33 | 21 | 0.00000 | 21 | 30 | 0.00000 |
| Grif '18, psilo (1 mg) + standard M/S practice | 25 | 20.1 | 20.5 | 0.00000 | 13.9 | 17.5 | 0.00000 | 30 | 18 | 0.00000 | 22.3 | 20.5 | 0.00000 |
| Carb '18, Placebo | 20 | 4.7 | 9 | 0.00000 | 6.5 | 9.3 | 0.00000 | 15.8 | 11.7 | 0.00000 | 6.3 | 9.8 | 0.00000 |
| Lie '17, MDMA (75mg) | 30 | 16 | 16.4 | 0.00000 | 5 | 32.8 | 0.00000 | 18 | 21.9 | 0.00000 | 9 | 27.3 | 0.00000 |
| Lie '17, Placebo (25µg LSD) | 4 | 3 | 0 | 0.00015 | 4 | 0 | 0.00018 | 9 | 8 | 0.00000 | 5 | 2 | 0.00001 |
| Lie '17, methylphenidate (40mg) | 30 | 7 | 16.4 | 0.00000 | 1 | 0 | 0.00000 | 11 | 21.9 | 0.00000 | 4 | 0 | 0.00000 |
| Lie '17, Placebo | 16 | 0 | 0 | 0.00000 | 1 | 0 | 0.00000 | 3 | 0 | 0.00000 | 2 | 0 | 0.00000 |
| Vlis '18, ketamine | 15 | 6.5 | 2.6 | 0.00000 | - | - | - | - | - | - | 6.6 | 2.1 | 0.00000 |

Table SM3: Identical values as to those Table SM2, except the formatting indicates statistically indistinguishable studies using a significance level of 1% ($p > 0.01$).

| | | Ineffability | | | Mystical | | | Positive mood | | | Transcendence of Space & Time | | |
|---|---|---|---|---|---|---|---|---|---|---|---|---|---|
| | n | mean | std dev | p value | mean | std dev | p value | mean | std dev | p value | mean | std dev | p value |
| **Isness-D** | 58 | 57.2 | 26.6 | n/a | 49 | 22.4 | n/a | 63.3 | 18.7 | n/a | 59.5 | 21.6 | n/a |
| Isness-C | 57 | 63.9 | 20.4 | **0.13287** | 61.5 | 16.5 | 0.00092 | 72.5 | 11.4 | 0.00191 | 66 | 13.4 | **0.05547** |
| Bar '18, MeO-DMT | 20 | 88.7 | 12.6 | 0.00000 | 79.3 | 18.4 | 0.00000 | 88.7 | 11.7 | 0.00000 | 85.7 | 13.8 | 0.00000 |
| Grif '11, psilo (30mg) | 18 | 85 | 25 | 0.00019 | 73 | 25 | 0.00024 | 79 | 21 | 0.00344 | 80 | 25 | 0.00113 |
| Grif '18, psilo (20-30 mg) + high M/S practice | 25 | 76.3 | 19.5 | 0.00179 | 71.8 | 18 | 0.00002 | 79.8 | 19 | 0.00043 | 70.6 | 18.5 | **0.02796** |
| Grif '11, psilo (20mg) | 18 | 79 | 25 | 0.00291 | 67 | 21 | 0.00346 | 72 | 25 | **0.11682** | 69 | 30 | **0.14317** |
| Nich '18, psilo (42mg) | 12 | 81 | 26 | 0.00609 | 65 | 35 | **0.04646** | 72 | 28 | **0.18513** | 73 | 31 | **0.07303** |
| Carb '18, psilo (30mg) | 20 | 72 | 18.8 | **0.02456** | 61.3 | 21 | **0.03471** | 66.3 | 20 | **0.54511** | 59.8 | 17 | **0.95525** |
| Grif '18, psilo (20-30 mg) + standard M/S practice | 25 | 74.4 | 31.5 | **0.01250** | 60.5 | 36.5 | **0.08258** | 74.5 | 21.5 | **0.01908** | 66.6 | 28 | **0.21369** |
| Nich '18, psilo (31.5mg) | 11 | 78 | 26 | **0.01988** | 60 | 35 | **0.18009** | 64 | 29 | **0.91788** | 65 | 31 | **0.47436** |
| Grif '16, psilo (22-30mg) | 50 | 74.5 | 26 | 0.00093 | 59.6 | 29.8 | **0.03758** | 69.8 | 27 | **0.14453** | 62.1 | 23.9 | **0.55396** |
| Nich '18, psilo (21mg) | 10 | 73 | 27.2 | **0.08842** | 53 | 37 | **0.64050** | 68 | 28 | **0.49964** | 54 | 33 | **0.49635** |
| Carb '18, psilo (20mg) | 20 | 66.3 | 23.3 | **0.17803** | 48.5 | 28 | **0.93597** | 60.5 | 24 | **0.59372** | 51.5 | 21 | **0.15449** |
| Grif '11, psilo (10mg) | 18 | 66 | 25 | **0.21782** | 48 | 25 | **0.87255** | 63 | 25 | **0.95651** | 47 | 25 | **0.04235** |
| Lie '17, LSD (200µg) | 11 | 49 | 26.5 | **0.35166** | 44 | 26.5 | **0.51192** | 58 | 26.5 | **0.42453** | 48 | 23.2 | **0.11415** |
| Grif '11, psilo (5mg) | 18 | 57 | 30 | **0.97850** | 43 | 21 | **0.31728** | 55 | 30 | **0.16278** | 44 | 30 | **0.01823** |
| Lie '17, LSD (200µg)-A | 16 | 83 | 12 | 0.00034 | 40 | 24 | **0.16539** | 65 | 16 | **0.74137** | 60 | 20 | **0.93391** |
| Carb '18, psilo (10mg) | 20 | 45.7 | 19.4 | **0.08002** | 34.8 | 19.8 | **0.01404** | 49.3 | 18 | 0.00468 | 35.2 | 16.2 | 0.00002 |
| Carb '18, dextromethorphan | 20 | 59 | 18 | **0.77973** | 29.7 | 21 | 0.00117 | 46.3 | 22.1 | 0.00128 | 49 | 21.2 | **0.06349** |
| Grif '16, psilo (1-3mg) | 50 | 30.8 | 31.7 | 0.00001 | 24.3 | 27.1 | 0.00000 | 35.8 | 28.3 | 0.00000 | 22.4 | 20.5 | 0.00000 |
| Grif '11, Placebo (0mg psilo) | 18 | 23 | 21 | 0.00000 | 19 | 21 | 0.00000 | 33 | 21 | 0.00000 | 21 | 30 | 0.00000 |
| Grif '18, psilo (1 mg) + standard M/S practice | 25 | 20.1 | 20.5 | 0.00000 | 13.9 | 17.5 | 0.00000 | 30 | 18 | 0.00000 | 22.3 | 20.5 | 0.00000 |
| Carb '18, Placebo | 20 | 4.7 | 9 | 0.00000 | 6.5 | 9.3 | 0.00000 | 15.8 | 11.7 | 0.00000 | 6.3 | 9.8 | 0.00000 |
| Lie '17, MDMA (75mg) | 30 | 16 | 16.4 | 0.00000 | 5 | 32.8 | 0.00000 | 18 | 21.9 | 0.00000 | 9 | 27.3 | 0.00000 |
| Lie '17, Placebo (25µg LSD) | 4 | 3 | 0 | 0.00015 | 4 | 0 | 0.00018 | 9 | 8 | 0.00000 | 5 | 2 | 0.00001 |
| Lie '17, methylphenidate (40mg) | 30 | 7 | 16.4 | 0.00000 | 1 | 0 | 0.00000 | 11 | 21.9 | 0.00000 | 4 | 0 | 0.00000 |
| Lie '17, Placebo | 16 | 0 | 0 | 0.00000 | 1 | 0 | 0.00000 | 3 | 0 | 0.00000 | 2 | 0 | 0.00000 |
| Vlis '18, ketamine | 15 | 6.5 | 2.6 | 0.00000 | - | - | - | - | - | - | 6.6 | 2.1 | 0.00000 |

**Table SM4**: The *p* values returned from the 'DistributionFitTest' command in Mathematica on the **I**, **M**, **P** and **T** averages for each participant, where the *p* value indicates whether the data are characteristic of a normal distribution. The data were categorized as being normally distribution if the *p* value was above a statistical significance level of 5% ($p > 0.05$). The **M**, **P** and **T** factor scores pass every test, and the **I** factor score passed all except the Pearson $\chi^2$ test (in bold).

|  | **Ineffability** | **Mystical** | **Positive Emotions** | **Transcendence of Space & Time** |
|---|---|---|---|---|
| Anderson-Darling | 0.12434 | 0.61534 | 0.07552 | 0.53278 |
| Baringhaus-Henze | 0.45081 | 0.56455 | 0.20420 | 0.90626 |
| Cramér-von Mises | 0.16477 | 0.55253 | 0.09207 | 0.49145 |
| Jarque-Bera ALM | 0.31246 | 0.58568 | 0.48709 | 0.59613 |
| Mardia Combined | 0.31246 | 0.58568 | 0.48709 | 0.59613 |
| Mardia Kurtosis | 0.30232 | 0.38436 | 0.35291 | 0.33066 |
| Mardia Skewness | 0.38654 | 0.63793 | 0.54530 | 0.90848 |
| Pearson $\chi^2$ | **0.03311** | 0.80075 | 0.12823 | 0.39732 |
| Shapiro-Wilk | 0.07108 | 0.57495 | 0.17330 | 0.45009 |

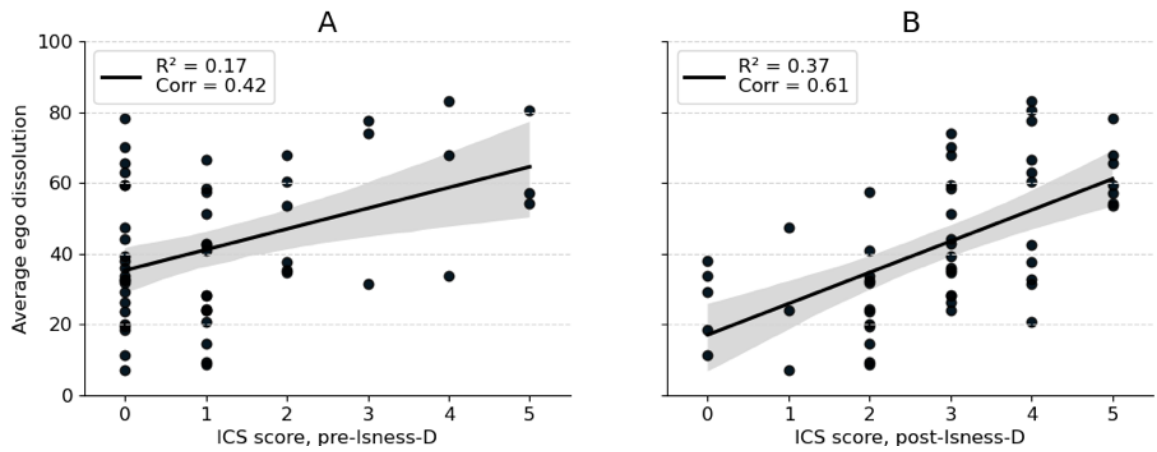

**Figure SM1:** Scatter plots of the average ego dissolution against the pre-Isness-D ICS scores for each participant (panel A) and the post-Isness-D ICS scores for each participant (panel B), with a line of best fit calculated using a linear least-squares regression. Each legend gives the correlation coefficient and the $R^2$ value.

| Item number | Item | Average | Std. dev. |
|---|---|---|---|
| 1 | During the ceremony, I felt a bond with my fellow participants that I could not experience outside the ceremony. | 5.2 | 1.6 |
| 2 | During the ceremony, I felt a sense of belonging with the other participants. | 5.5 | 1.2 |
| 3 | During the ceremony, I felt a sense of harmony with the others. | 5.9 | 1.1 |
| 4 | During the ceremony, I felt a sense of sharing with the other participants. | 6.0 | 1.1 |
| 5 | The ceremony really allowed me to get to know the other participants. | 3.6 | 1.4 |
| 6 | During the ceremony, I felt that social status became irrelevant. | 6.4 | 1.0 |
| 7 | During the ceremony, I experienced all participants as equal based on basic shared humanity. | 6.2 | 1.3 |
| 8 | During the ceremony, ego-related tensions between participants seemed to resolve. | 5.3 | 1.4 |
| 9 | During the ceremony, I felt a strong connection to another participant. | 4.3 | 1.9 |
| 10 | During the ceremony, I felt a strong connection the facilitator. | 4.6 | 1.5 |

**Table SM5:** Mean rating and standard deviation for the 8 items in the Communitas questionnaire (items 1–8), and two additional items relating to connection with another participant and with the facilitator (items 9 & 10).

Table SM6: Themes that arose in the qualitative analysis and several corresponding indicative quotations, in order of the number of associated quotations.

| Theme (N statements) | Indicative quotations |
|---|---|
| **Connectedness (75)** | - I'm pretty bewildered… It's very immersive… you lose yourself in the connection to other people.<br>- It was the antithesis of what we're going through right now, with all the separation.<br>- *Blissful sense of connectivity.*<br>- You can have this quite intense… intimacy in space… but you are also totally safe. You can be with strangers and… be really close but also know where your boundaries are. |
| **Positive emotions (65)** | - I feel remarkably happy. There's something very happiness-inducing about this whole practice.<br>- It was quite a trip. But a pleasant one.<br>- For me that was very very calming, amazingly calming.<br>- I feel way less stressed. |
| **Embodied awareness (40)** | - Even though there was no body there, you wanted to respect each other's space.<br>- It felt really nice to touch people. Having not touched people for so long [in lockdown], it felt really nice.<br>- *Following the [dynamical molecular organism] with my nose, it felt like I was breathing it in.* |
| **Ego dissolution (31)** | - Identity didn't matter anymore; it was about experiencing things together. That was wonderful.<br>- I found [the anonymity] quite powerful because we were all completely equal in the space. Any of the pre-judgements that come in from how people look, sound and that sort of thing just aren't there.<br>- Yeah, I felt [when touching one another's heart centres] the sweetest tenderness or pure, childlike love… stripped back, without any of the assumed layerings that we place upon reality and relationships… just to the absolute core, it was truly beautiful. Thank you, all of you. Thank you, Isness. |
| **Supportive setting (28)** | - At first, I thought [it] might be difficult to [reflect on connection with] strangers, but there was no pressure to give a long statement… the breathing exercise helped get comfortable in the space.<br>- *I really liked the opening and closing exercises with [the facilitator]. It really helped to hold the experience and make me feel more comfortable with everyone.*<br>- *The time of reflection at the end of the session, hearing other people's thoughts on the experience made the experience more meaningful.* |
| **Sense of play (24)** | - There certainly was a sense of playfulness in this, which was really nice. As adults that's not always something that we pursue. It was good.<br>- I was inspired to dance on the top of the table at one point.<br>- *A feeling of being grounded, a playfulness facilitated in a calm way.* |
| **Transcendence of space/time (13)** | - It felt like the space opened up to me. I'd been aware of the confines of [the space], vaguely. And then we pushed [the molecular organism] through the floor… it was very odd to feel like everything opened out.<br>- Things didn't have the same human scale, so at times it felt this could be [an enormous] space and we were a completely different scale, or it could be microscopic, and it didn't really matter, actually.<br>- *I liked the parts that allowed me to view the world in a way that isn't possible in real life.* |
| **Noetic quality (11)** | - I got quite emotional… And then the experience was guiding us forward, so it was fleeting. But I got this surge of emotion where I don't know if I wanted to gasp or cry or what it was, but I was kind of shocked in awe.<br>- That's definitely a spiritual experience of some sort. It's, like, tangible.<br>- *Pure unity.* |
| **Comparison to other altered states (10)** | - [Taking psychedelics] wasn't quite like this, but I guess there's elements of it, for sure.<br>- *I experienced the sense of interconnectedness…only previously noticed with the help of psychedelics in the right setting.*<br>- *The experience brought up memories for me of medicine plant journeys and realisations that have shaped my life and give my life meaning. These realisations are very much connected with going beyond form…* |
| **Sense of beauty (9)** | - I was amazed [by] how moved I was. I think it was the juxtaposition of the frustration and then the beholding beauty. I think the music really helped. It felt very beautiful to sit back and witness. I don't normally get moved quite so much.<br>- Yes, thank you. I was a little bit doubtful in the beginning, I didn't know what to expect and what was going to happen. It was a beautiful experience. |
| **Reflection on mortality (8)** | - It's clear that [our life] energy isn't just gone, it goes somewhere. It's beautiful to think that it's all out there somewhere and it's still circling you constantly<br>- There was one moment… talking about seeing past energies and I was thinking about my cat, whom I lost last year. I really miss her. I could really see her and… was writing her name. It was beautiful, thank you. |

**THEMATIC ANALYSIS**

A summary of the identified themes (N quotes):
1. Connectedness (75)
    a. General (42)
    b. Through body-movement (19)
    c. Intimacy (14)
2. Positive emotions (65)
    a. General (40)
    b. Feeling of calmness (25)
3. Embodied awareness (40)
4. Ego dissolution (31)
5. Supportive setting (28)
6. Sense of play (24)
7. Transcendence of space and time (13)
8. Noetic quality (11)
9. Comparison to other altered states (10)
10. Sense of beauty (9)
11. Reflection on mortality (8)

Each quotation has been formatted according to the source:
- Regular = Guided discussions
- **Bold = free writing responses**
- *Italic = questionnaire feedback from "what did you like?"*
- ***Italic bold - questionnaire feedback from "Do you have any other comments?"***

**Theme: Connectedness**

*General connection*

1. At first I was a bit suspicious with this network, but I feel that I'm starting to understand the metaphor a bit better...The one thing that I was a bit wary of was the fact that we had to go about and do the mudra in order to interact with it. I thought at first it would go against the idea of being always connected, but I guess now I feel like in order to be connected we have to put in effort, we have to understand that. Being connected is not something that should be expected all the time. And it takes some effort to realise it, but now I feel like, I don't know if that's going to be a changing experience, but now I'm ready to have this metaphor in my head whenever I'm scouring through life.
2. Yeah, when we were doing the initial mudra and [the narrator] said change to the other one and see what you notice. But that initial one, yeah, I could just get a real connection with [the molecule].
3. I feel very calm and centred...It's really nice. Connected as well, you know.
4. I've been also trying to be more connected with myself through meditation, and it felt really nice to have a meditation with other people. So, I felt connected with myself but also with everyone else here. It's an experience I haven't felt in a while because of COVID and being in front of screens all the time. That's why I think 'connected' is the word for me for the end of this session.
5. This is the closest [that] I've been to any strangers in a year.
6. It was beautiful. It was amazing. I kind of feel like I know you guys now.
7. It felt more useful [than other guided meditation], in a sense a bit more connected.
8. I think it's nice that we're all in the middle together now. We all started out on the outside and we're all together now, it's nice.
9. I suppose it's a bit more collaborative than just a normal meditation because we were able to work together and connect with each other in different ways.
10. I'm pretty bewildered, to be honest. It's very immersive. It feels you lose yourself in the connection to other people. Quite remarkable.
11. Yeah, I feel grateful as well, for being so connected to the sense beyond touch and taste and smell. Very clear connection in a way [trails off] beautifully connected in this experience.
12. It was also just so amazing to feel that movement in myself to connect, and that readiness. All I wanted to do was to be together. Such enjoyment of being together and to feel that so powerfully in oneself, that is really confirming and opening and expanding.
13. Such a pure joy of connection.
14. Connection for sure, yes, and so much more.
15. [When the lights faded] Still connected.
16. Especially in the situation we're in at the moment, globally, not only was there, on a practical level, more freedom to move and [no] fear that you were going to whack someone in the face. But also, you could physically get closer to one another's lights, which was really cool. You could be in the light, which you couldn't before. But also, the connection is so much more than who you're physically in the same room as, it's about a bigger thing, feeling connected to people we're not physically connected to, necessarily.
17. I wish it was bigger, I don't know if that makes sense. That there was more space and, also, more of us...It felt like at any moment the lights could come on and there would be 6 billion more.

18. I think the fact that we're all doing it together makes such a difference…It's the fact that you realise that it is someone else, an actual real person making those moves.
19. It was the antithesis of what we're going through right now, with all the separation.
20. It was weirdly connected [when coalescing].
21. **Connecting light and emotions as energy was special. My work is in renewable energy and I also take photographs and think about light; light as connection; light transferring between matter; light creating memory. Experiencing myself and the other people in the group as light energy was joyful. It was both another space and allowed me to think about other spaces connecting in the world.**
22. **Loved the thought of connection with total strangers, in such a gentle way. Thank you for the experience!**
23. **It was really nice to feel really connected to people – especially during COVID to feel so free to touch & connect with other people. Felt very strange standing inside someone, but also lovely.**
24. **I noticed that I enjoyed the ability to join with people and leave when I wanted — togetherness and separateness. The option does not always seem available to choose for myself. The music was a significant part of the experience for me — relaxing and connecting. The attention to a separate 'self' was intriguing. The energy part. The energy among the lights, (people) was easier for me to experience — I experienced some type of connection.**
25. **I liked the fogging effects that connected us all in haze. Very soothing and centering experience overall. I was sort of expecting it to go from protein folding to a tree growing. I liked us being amorphous jelly fish together. Feeling connected to others —distance irrelevant.**
26. **Actually, this great 'therapy' in time of social distancing. Happy to participate again and again.**
27. **Connectiveness — good moment of release.**
28. **Wished to see and feel all the people on Earth in the experience, since the idea is that we are all light beings. Would be incredible to experience that directly.**
29. **I felt at ease with the other participants and felt a positive connection with them at the end.**
30. *Blissful sense of connectivity.*
31. *Everything! I liked being energetically Together yet Physically apart from the other participants a lot. It increased the sense of connection.*
32. *I felt excited by the international connections and the use of the Internet as a way of creating relationships between people whom I have never met. I felt a sense of joy and connection which, in part due to lock down, I have been unable to experience as my normal dance practice of contact improvisation has been suspended.*
33. *That I was able to feel immersed in the experience. Feeling of connection with a group of strangers who were spread out across the country.*
34. *The potential for bringing people together.*
35. *Connecting with the other participants, interacting with beings as balls of light.*
36. *The opportunity to connect with other beings and myself in a new way.*
37. *The feeling of calm and peace I felt and the thought that I was connecting with strangers from around the World going through the same experience.*
38. *I like that I have a visual now that I can map onto the real world when I'm around other people so that I can see everyone as that beam of light.*
39. ***I wish I could do it again, now I know what's involved. It was so beautiful; I felt an amazing connection to the lights and energy. Now I think about it, maybe I did feel more 'at one' than I scored in the questionnaire.***

40. *Really great, I would love to do this with my siblings across the world as I feel it would make me feel so close.*
41. *Really beautiful. My intention at the start was to look more at my connection with myself and my body and I really feel like I was able to explore this from a new and different point of view. Thanks guys!*
42. *I think I preferred the remote experience more than when we were in the same room. It felt more profound and in line with the project. I also wasn't afraid I was going to whack someone and so I could move more freely. It also meant we could get closer than before which felt more intimate, and connecting—nearly as much so as with a partner, child or pet—even though we were in different places.*

*Connection through body movement*

1. I think I was still aware that I wanted to help out what everyone else was doing, rather than ruining it, if that was such a thing. If someone was trying to go a certain direction, then I wanted to go with them rather than spoil their activity.
2. Today felt so much more like a dance floor. I guess you feel that thing when you're on a dance floor in a club with your mates and you feel that sense of connection. And you're moving in a similar, exploratory way. Today, for some reason, [Isness] felt more like a disco.
3. I just wanted to move in, to be honest. I was just really curious to see what happens when you put more and more light together. And I think people were receptive, I don't know if that's just in my head. When I was trying to get as much light together as possible, I felt like a lot of people were following those same thought patterns. And you definitely felt that when we were all bunched up on top of each other, that was really cool.
4. I felt really connected with all the other people in this room, even though we're in different places. It was really nice all putting our lights in the middle, it was a way of feeling connected.
5. It felt like there was a way to tell whether people wanted to [coalesce] or not, people kind of moved gently and moved direction.
6. Naturally, I wanted to start moving slower. And I felt like the interaction could be greater perhaps when we were moving slower. So, I feel like I slowed down during the process. I started a bit more like 'woah, what's going on', and then I was enjoying slowing down and observing others and seeing how others were interacting if I went near them.
7. I also found it really cool as a social experiment, to see how leading and following happens within the group before it started being led. When we were just left to our own devices, what we implicitly did with the creature—trying to stretch it out—came from someone and everybody else just picked it up. How quickly that happened was fascinating.
8. **Really enjoyed that. Loved feeling of being connected with people — sense of a very intimate space despite being so far apart (one in London and one in Knowle, Bristol). Lovely working together. You get a sense of whether others in the group are playful, generous, co-operative, from how they interact with the shapes and the other people. So — body language still comes through even without a physical body.**
9. **OK, trying not to think too much but I can't help it. I'm really into the idea of building the sense of interconnectedness in this way. Simple; three moving lights to represent a stranger. Was enough to really feel and play with a sense of connection. There was mirroring, suggesting/influencing, and being influenced. Plus a shared novel experience.**

10. **Social experiment – play and unspoken coordination. I really liked how quickly unspoken coordination to stretch out the creature developed – way before we got 'the directive'. The directive then felt really out of touch.**
11. **I didn't really connect with the 'beings of light' script. Instead, I thought the participants convinced me they are good, well-intentioned, and bright people ('that carry light') all on their own, through interaction. The whole recorded script fell flat in the light of this realization that was about reality, not some pre-recorded open-loop non-judgement.**
12. **Doing simple actions, having to go into the middle, choosing a movement helped put me at ease. Guide stating to be gentle with the molecule set up the conditions of we should interact in the space – 'be gentle'. And each other. There was a sense of play which you don't have in the real world because of social constraints. I felt everyone was respectful of each other's space, no one was dominant, more curious and wanting to gently interact.**
13. *I tend to be quite shy but felt I got easily and quickly comfortable with the others, through body movements and interactions mainly.*
14. *I loved playing with the energy and manipulating it. It was great when we could all coordinate our movements.*
15. *I enjoyed playing particularly with one of the participants, improvising and guessing each other's moves.*
16. *That there were a small number of participants, so we could interact in a small group. Everyone was friendly and playful. The visuals, especially the trails and fog effects were cool. Working together to stretch the protein/organism/thing out was fun.*
17. *The way it became a dance.*
18. *I felt free to dance with others and play freely.*
19. *I feel much closer to my intention set at the start than I did before the session. I enjoyed the live guiding and group engagement a lot and felt very safe and connected. The pre-recorded guide sometimes seemed too frequent or didactic so there wasn't as much space as I would have liked to play, explore, learn from watching others through chance encounters.*

*Connection in intimacy*

1. It feels really much more intimate, so much more intimate [than a video conference].
2. The fact that you can have this quite intense, almost, intimacy in space, you are right in each other's space, but you are also totally safe. You can be with strangers and you can be really close but also know where your boundaries are in certain ways.
3. It felt really intimate. And I tried to connect whichever light being it was, and you didn't know who it was. I tried to join mudras, and I think I sat inside a few of you. I hope that was OK!
4. It's also really interesting because the first time I tried this there were four people in the room together, so we ended up bumping into each other and you had to be aware of space and cables. And now, I can just literally walk into people and it's quite sensuous, almost.
5. It felt strangely intimate...it was really nice.
6. **Feel slowed down now. Liked connecting with others. More intimate than would have thought. Enjoyed the visuals of my own movements, was less interested in others.**
7. **Overlapping in 'virtual space' almost felt uncomfortably intimate at time—probably a good exercise!**

8. **Physical boundaries being breached was initially really jarring but quickly gave way to feeling of intimacy. This was a really good part of the experience.**
9. **Super fast intimacy and trust establishment. Following and leading, playing one on one led to a feeling of true intimacy, which was a surprise. Was completely emergent. I used to do improv – this kind of exercise would establish physical trust much faster than happened in real life.**
10. **I found it incredibly energizing. Crawling through the light, the creature we were all holding, as well as the [?] light of other beings. It was sensual and playful and safe, felt like feeding off each other's energy. Also, quite erotic.**
11. *The sense of an intimate space despite being far apart.*
12. *The closeness, intimacy, body-meld.*
13. *The sense of intimacy.*
14. *I thought the playful intimacy did come about very quickly, I have experienced real life playful intimacy through Contact Dance to the same kind of depth however, I would agree and echo one of the other participants that this happened very quickly.*

**Theme: Positive Emotions**

*General*

1. This was amazing.
2. And just I guess a feeling right now, not that it was part of the experience, but just a great conversation with a bunch of glowing gas balls. Absolutely loving this, it's something I never knew I needed.
3. Just a whole lot of gratitude, [Facilitator's name]. This has been absolutely gorgeous.
4. Yes, thank you so much for this opportunity, I never felt like this before. And thank you for that.
5. When [the molecule] was leaving behind trails, I found that particularly amazing. That really felt extraordinary.
6. Yeah, [coalescing] feels a bit funny at first, quite jaunting almost. But then it feels quite nice, like 'ah yeah, OK!'
7. I feel like there's a difference between stagnant and stillness. That's how I'm feeling. That it's nice to know the difference. I guess this time can make you feel very stagnant, cause you're in the same place and you're not still or rooted to anything...I feel like there's a difference.
8. I liked [coalescing], I thought that was really nice.
9. [Coalescing] was great.
10. [Coalescing was] a quick way of freeing you up and making you realise that you can walk across other people's space.
11. I was expecting to walk into you guys, but [we were] disincorporated beings. It was great.
12. [Coalescing] was a highlight, an absolute highlight. It was very interesting having done the physical version of this and seeing the advantages of the virtual. That was definitely one of them, being able to invade and conquer someone else's space.
13. I like silence in the space. It's really interesting the experience of silence in this dark space without anything at all. Not only the experience of light, but [also] the experience of silence and darkness is really great.
14. Yeah, I was going to say that I haven't really got any words. Which is a good thing, I think.
15. [Coalescing] was definitely a highlight.
16. [Coalescing] was really cool.
17. I felt quite greedy for it [the molecule]. I just wanted to be in it and it to be completely surrounding me and to be amongst it. I still feel that.
18. I feel remarkably happy. There's something very happiness-inducing about this whole practice.
19. I feel happy and pretty optimistic about going out and seeing energy in everything and having that as a positive lens about life.
20. It was quite a trip. But a pleasant one.
21. **I'd like to do again in the morning, though curious what I'll be in dream state tonight.**
22. **I really like that the creature is common to all.**
23. **Felt a strong urge to disappear/let go of mudras. When I did, I felt elated and giggly.**
24. **Overall, very impressed. Thanks to everyone who worked on this, it's going to be a beautiful, revolutionary public service.**
25. **The previous time I experienced this, there were people actually in the room with me and I was conscious of their personal space, headset cables, etc. Now it was so much more freeing and … safe?**
26. **Initially I was thinking about driving on the motorway with other death machines around me, which makes me anxious. Translating that sense of danger in a completely safe space, yet with other live participants who you can still FEEL almost physically, I was delighted that it didn't diminish the sense of sharing energies at all. That is amazing!**
27. **I feel so [?], awake, glowing, even after a long day. Transcending time, time zones, other humans in the experience with me … I feel so full of LIGHT and HAPPY.**
28. *All of it.*
29. *All of it!*
30. *It was a really GORGEOUS experience... this iteration is more powerful than the Church one I did 8-9 months ago…the music adds a lot…*
31. *Great experience, thank you.*

32. *It was very centering!*
33. *Brilliant.*
34. *I really enjoyed it, and it is something that I would definitely like to do again in the future.*
35. *No. It was fantastic! :)*
36. *Thank you. I'm humbled and grateful to have been able to enter this space. It arose so many things from within and beside me. Thank you! Very fun and very humbling.*
37. *I had a great time, but mostly forgot my 'connectedness intention' that I had been asked to think about before the isness experience as I played with the energy string and the other participants. I was aware from conversations with others about the inspiration from chemical modelling so that may have been in my mind when I did that.*
   *It is something I think I will remember for a long time, and would love to have another play!*
38. *I never thought this could be done with VR. Amazing. Please keep going.*
39. *Many thanks! Amazing :)*
40. *I really enjoyed it!*

*Feeling of Calmness*

41. Tranquil.
42. Yes, absolutely [I feel chilled]. I'm like a lukewarm glass of water.
43. [When the lights faded] Like going to sleep as a kid. Maybe at the end of your birthday, or something. Or falling asleep in the car when your parents are driving.
44. [When the lights faded] Like when you're little and you fall asleep and your parents are having a party. You know there's those times where you fall asleep on the sofa and could hear other voices around you and that sense of togetherness.
45. And I definitely want to do it again. Now I've experienced it, I want to do it all over again and I can relax into it a bit more. I do feel relaxed now. Especially that point where our lights all got on top of each other. It was relaxing.
46. I feel way less stressed.
47. Yes, very relaxed. Really calm. And I was amazed at how I wanted to gather all the mudras together. I wanted to bring everyone [trails off]. I didn't like to see people straying off.
48. I felt very peaceful, I think.
49. For me that was very very calming, amazingly calming.
50. I will say that I'm relaxed in a way that I can describe, I think that my shoulders are very much lower, I had a tense neck. And I loved the dance, I was dancing with my hands.
51. **Overwhelming calmness once I had let myself fully immerse in the journey!**
52. **The experience was fun and eye opening! Felt very meditative and calming. It was cool interacting with the 'organism' as well as the fellow users.**
53. **Overall, a very joy-inducing and calming experience.**
54. **My first VR with headset. Reminiscent of Tokyo Light Digital Museum. Very relaxing (recharging).**
55. **All in all, I felt renewed/re-energized. [?] Would be happy to do again.**
56. **Heart opening.**
57. **Great for group therapy. Great for group tonglen practice.**
58. **There was a point when you were asked to focus on the past as you looked at the light trails. I had a spike of an anxious feeling – perhaps because depression is an obsession with the past…and I have always looked at it in a way where I focus on the negative. What was interesting was I didn't feel like that; I focused on the non-linear patterns. It was nice to look at the past in a calming way.**
59. **I found it difficult to write about this experience straight afterwards. Although I felt lucid and wanted to talk to my host about it, I have had very limited experience with VR and was focused on feeling calmer and strangely energized from the experience.**
60. **Some of the questions in the survey were difficult to answer because I didn't come into the experience with an ego – my starting place was a feeling of anxiety and stagnation. I came away from the**

**experience with an overwhelming sense of calm, which onto the next morning. I woke up thinking about how I keep a sense of play and wonderment.**
61. *Interacting with the energy string was fascinating and soothing.*
62. *The use of VR to aid the disillusionment of self as a starting part to a meditative practise.*
63. *I liked the immersive nature which removed me from my normal thoughts and experiences.*
64. *Calming, relaxing feeling at end.*
65. ***I'm not sure I took away any profound insights, but this may be because I have never really tried yoga or meditation, which would perhaps make me more primed and open for this type of experience. It was, however, quite relaxing and I would try it again.***

**Theme: Embodied awareness**

1. I almost…we were different creatures. In fact, I remember now the dance movements were these fuzzy, round...you know...we are these fuzzy round cloudy things, and that was a great form for the particular dance.
2. It felt like we were all, you know, coming together, like I could *feel* that. That's what it felt like, I think.
3. Having done mediation without visual aids, I've not really done group meditation before. It's a similar process although I definitely feel an awareness of other people in my space. There's a duality in the fact that you're not in my physical space. Definitely, I think that at the end when I closed my eyes, I could still feel things moving around me. It's hard to separate traces that are left on my visual senses, differentiation the physical from the mental.
4. I thought it was very interesting to experience the same space without bumping into anybody. And just becoming into the same space with other people and merging, sort of, and at first I feel a bit 'OK I'm going to bump into them' but no, I can completely merge with them, it's wonderful.
5. Even now, I'm arranging myself sat down and I'm like 'ooh have I given space to the person on the left and the person on the right' and it's kind of incredible.
6. There's definitely a feeling here like 'someone's messing with my lights' [and] 'somebody's standing in my orb' or 'somebody's standing in my space' or 'am I going to touch someone else?'. Like 'is that allowed?' [or] 'is that taboo?'.
7. It felt really nice to touch people. Especially in lockdown, having not touched people for so long, it felt really nice.
8. I think I still found myself respecting peoples' personal space; it felt weird to try and walk over someone or walk through someone. I did treat people as if they had a slight solidness to them.
9. I found myself putting my ball of light on other peoples' hands and then feeling self-conscious that I was an inappropriate guy at the party.
10. I totally stepped into all of you purposely, I wanted to see how that felt like. I was trying to put my head on your other heads and my hands in the same spots, just to see if that made me feel anything. It felt like someone else's energy.
11. I always find it interesting when you give people more space to these, let's say, 'imagined people' than you would in real life. You can sense an imaginary presence around the glowing light, and you give it space even though there is no barrier there, other than imagined. But, with people in real life, even though it is, let's say, 'real' or the barrier is real, you push that more so than you would here, where you give people more space.
12. Even though there was no body there, you wanted to respect each other's space.
13. I wasn't sure if something else was going to happen when we did that [coalescing]. I don't know what I was expecting. It was weird, I suppose, not touching because it feels like we should have been, if you see what I mean.
14. I think the 'not touching' is part of starting the dissolution of the body and translation into this kind-of more 'energy state', and so it felt, when you couldn't touch someone, you were like 'that makes sense' because you're no longer a body, but you're light.
15. I found it really strange at the start when you were telling us to go closer and closer and then that weird boundary of where personal space is...and all of a sudden it was like 'oh, actually this is kind of nice!'. Like, we're all in the same block together.
16. Yeah, [coalescing] was really nice.

17. I felt aware of that, I think, more so than I needed to, being in someone's space.
18. I think part of me, I don't know, there was something in me that felt a little bit rude walking into someone else's space, despite the fact that I'm not actually doing that. It felt quite intimate with strangers.
19. I kept wanting to touch people, physically. To reach out and grab their hands or hug them, see what would happen if our energy came together.
20. I actually had a sensation of different temperature on my skin. And my heart opened. It was definitely heart-opening. Much more rapid, I think, from the regular loving, kindness meditation, even. It speeded up the heart-opening for me a little bit. Yes, I felt really deeply connected to all of you. In a human but non-human way.
21. When we could then, right at the start, we all moved closer and closer together and we [were] on top of each other's lights. Until that point it really felt like you were all in the room with me, in a sense, and then that made me realise that you weren't. I thought that was really interesting.
22. I kept being concerned about people's personal space and not stepping on them.
23. I was a little surprised at how connected I felt with the other people. I think, like the other gentleman was saying, somebody coughed, and I thought 'Am I too close to this person?' It was kind of strange.
24. In one way, I felt like there wasn't enough space. On the other hand, the constricted space, sort of, increases interactions and like, yeah, there was something happening. That was an interesting dichotomy for me.
25. What was interesting for me [was that] we couldn't see each other's body forms. At one stage I was lying down and at one stage I was kneeling, and I thought that was interesting. I was curious what everyone else was doing.
26. I was struck by how quite quickly something that was quite abstract—the light—suddenly held a lot of meaning when you perceived [them] as being people. I was quite struck when we all walked into the middle together, how that felt really real.
27. [?] get the proprioception, you know, to figure out how far you were from the other person and all that. But it was kind of fun, that was fun.
28. **Expected haptic feedback when connecting with creature and others, then felt it anyway, a little.**
29. **I enjoyed the subtle movements of the gloves. You felt aware of your body, but mostly more focused on the molecule.**
30. *It also felt really cool that I felt the heat from the light.*
31. *Whenever I would physically merge with any of the other participants, I got chills.*
32. *I loved touching people—I think because of the lockdown this enjoyment was even more heightened.*
33. *It also felt really cool to actually step inside others and also put [the molecule] in the ground.*
34. *Following the [dynamical molecular organism] with my nose, it felt like I was breathing it in.*
35. *When we were smoke.*
36. *I really enjoyed when we climbed on top of each other and all stood in the same spot.*
37. *I was struck with how quickly the abstract lights grew to hold tangible meaning, to understand them as other people. It was especially poignant when we moved toward the centre and felt as if we might collide or enter each other's personal space.*
38. *When bodies came together it felt like I was being tickled.*
39. *Being able to get closer to people and become the same light. Knowing that we were all in different places having a shared experience that wasn't on Zoom.*

40. *I was not sure if there was an effect when I touched or if someone else touched my heart light. In some ways I felt that I would have liked to be able to show my consent towards someone else interacting with my heart light.*

**Theme: Ego Dissolution**

1. For me, the whole feeling of just being in that space was that my identity felt lost. And then when you expand you feel like the [vastness] of the space that you're in and the energy that forms meant [for me] that myself as an identity didn't matter anymore and that was actually really liberating.
2. Yeah, it was very beautiful. I really loved how it challenges the relations. Like remember when we just gathered around, the worry at first [that we] would hurt one another, and then it quickly vanished. Now I just feel like it helps you dissociate from something within you. I think it can relate to what you said [participant name], about how your identity gets a bit lost. And I don't know if it for me was my identity or my anxieties that I put on the side. I feel present right now.
3. Identity didn't matter anymore; it was about experiencing things together. That was wonderful. There was an exterior manifestation of identity, yeah, it was new.
4. It was more like a space to be free, you could do whatever you want and not feel judged. A bit different from going to the gym.
5. I'm really enjoying [the current moment] when somebody talks, turning to them as though you're going to look in their face but you're just looking at this shape. And that feels really nice! Knowing that's a person and hearing your voice, and everything's live and real and connected, but then there is also this really abstract element. That's really fun.
6. I enjoyed not knowing who was who.
7. [The anonymity] Relaxing.
8. I feel like when you meet people outside of here there's so much superfluous stuff on top; there's layers of personality and layers of expectation that they also portray. Weirdly, here, when you're seeing someone just as a blob of light and two little blobs, which you can assume are hands, you lose all of that and you see more of a personality. I know that sounds weird. I almost see authentically. Like a childishness, like a pureness in it.
9. Takes down the ego a bit. I just mean once you're talking to a point of light, I think it produces that sense of the other person's inner sense and yours and, I guess, would allow you to talk in a more matter-of-fact way.
10. I found [the anonymity] quite powerful because we were all completely equal in the space. Any of the pre-judgements that come in from how people look, sound and that sort of thing just aren't there.
11. Yeah, I felt that also [when touching one another's heart centres] and, the sweetest tenderness or pure, childlike love in how stripped back, without any of the assumed layerings that we place upon reality and relationships and just, to the absolute core, it was truly beautiful. Thank you, all of you. Thank you, Isness.
12. It's a completely other way of connecting that I'm not familiar with. All the usual stuff disappears.
13. There was a freedom in being there.
14. Yeah, I think so. The fact that you can't see anybody and it's just you [trails off]. You can't really be self-conscious.
15. I felt a little self-conscious at the beginning when everyone's position was established but once everyone had moved around, I felt much less self-conscious.
16. **Everyone together moving arms from side to side was magic. We were all the same. We were all one. We all looked the same but had our own spirits.**
17. **Group dynamics – rejection, not being part of it, missing the point. Helpful to realize as we were all equal this was completely my filters. Able to let those anxieties go. Judgments of others based on voice and what they sound — slowly melting away.**
18. **In a time where the human race is being told to be socially distant, it is fantastic to have this representation of how connected we are in reality, and that having to stay physically apart doesn't prevent us from mentally and energetically connecting to each other. Also, that sometimes what actually prevents deeper connection is the outward assumptions we make of one another based on physical appearances and social conventions.**
19. **The experience was strange. It seemed to reduce relationships between people to the simplest concept—connectedness—but without visual cues and no judgements that go with it.**
20. **It's a lot of easier to focus on people when they are just 3 points. I felt a bit more able to understand and interact in the absence of extra cues. Maybe that's just my [asperges] talking though.**
21. **Love the smoke effect and the ego erasure.**

22. *The sense of connection with the other participants at a real level not superficial, societal, etc.  Just them as they are.*
23. *That you could tell the personalities of people around you despite not being able to see them.*
24. *The sense that everyone was the same but with different energies and then the blending of these energies and the anonymisation of them as the session progressed.*
25. *I also liked the sandbox nature where we were anonymous and depicted only by points of light.*
26. *Feeling of anonymity, connection, playfulness, equality.*
27. *I enjoyed the sense of seeing others while not having to see the superficial form, which led to the felt sense of connection as beings.*
28. *The inability to perceive people as I normally would.*
29. *Loss of personal space.  Being able to connect with other people as if touching was particularly enjoyable due to current COVID-19 restrictions.  Meeting strangers as light forms with voices reduced preconceptions and helped us connect quickly, I think.*
30. ***Interacting with others' disembodied gestures had a quality of dissolving boundaries, gentleness, and warmth.***
31. ***It reminded me slightly of a trip, but mostly of a very intense and beautiful meditation session.  The one big difference being the shared aspect of it.  The connectedness of the group in the middle of a dissolved ego was really fascinating.  Thank you!!!***

**Theme: Supportive Setting**

1. I loved [the chaperone line]…At the beginning I didn't really pay much attention to it, but by the end I was like 'the blue line makes me feel safe and really contained', both in physical space but, maybe because of that reliance, without really thinking about it, it made me feel like it emotionally or something as well. By the end, I was like 'I love the blue line!'. And then I was surprised by that.
2. The talk that you gave, [Facilitator's name], at the start was really helpful as well. Especially the David Bohm quote, I was really interested in the general gist of 'the luminosity of this is not really available to us with our current brain chemistry'.
3. **Really nice to have little bit of getting to know [?] (movements/dancing) at the start to warm up and make everyone comfortable. [Facilitator's name] is a brilliant host!**
4. **The smoke-light point was particularly nice. I very much enjoyed the mudra lights and the organism.**
5. **Nice space. The touching the ground thing was cool. I liked picking the molecules.**
6. **Camp fire start/end was a nice touch.**
7. **Lovely narration, visuals well produced with further R&D could be even better.**
8. **[Host Name]'s incense gave a real presence to the body-smoke. That worked REALLY WELL. Do that. A mid-way breeze cleared the room of incense for a bit and it really changed the experience. I was glad when it returned.**
9. **I loved the bit where the 'history trails' came to the creature. Made it feel a lot less spastic.**
10. **[My host] put incense on at the moment smoke appeared. I found that really enhance the experience.**
11. **I've been in several group meditation scenarios (in person) and always found them invasive and stifling. I often respond badly to 'energy-policing' that is sometime found in such environments. The Isness platform is comfortable and safe in a way typical group meditation environments often aren't.**
12. **Setting an intention – at first, I thought that this might be difficult to [reflect on connection with] strangers, but there was no pressure to give a long statement. I felt the breathing exercise helped you get comfortable in the space.**
13. **The repetition of the woman's voice helped situate you within the space. Gave you a focus.**
14. **I thought the guide did an excellent job of setting the tone without giving too much instructions or being intrusive.**
15. I felt the guide was relaxed and not overpowering. [Facilitator's name] put me at ease.
16. I loved how gentle [Facilitator's name] was of a guide, the guided meditation script was on point, and the progression of activities and visuals were wonderful. I loved how in a matter of moments I felt a similar peak experience that I am more accustomed to experiencing with skilled dancers.
17. The guided meditation was super. Loved the woman's voice.
18. I really liked the opening and closing exercises with [Facilitator's name]. It really helped to hold the experience and make me feel more comfortable with everyone.
19. The reminders by the guide to be gentle with it because it held everyone's intentions was a lovely touch.
20. Loved the narration.
21. I enjoyed the quality of the guided meditation and the tone of voice of the facilitator.
22. The female narrator and overall experience was excellent, but I cannot say the best ever.
23. The sense of being present and absorbed by the experience. The music really helped with this. Great music!
24. The music and the gentle lighting.
25. The campfire start/end was a nice touch—it felt like other ceremonial experiences, and setting out on a journey.
26. The time of reflection at the end of the session, hearing other people's thoughts on the experience made the experience more meaningful.
27. Enjoyed the sharing of impressions with other participants in the end too.
28. Most of the visuals and audio were great. I really enjoyed the smoke effect, the spatial weirdness, and the interactions with the light creature.

**Theme: Sense of play**

1. I had a similar experience, with realising the point isn't about trying to control it, the point is just to enjoy it and play. Yeah, I really enjoyed it.
2. There certainly was a sense of playfulness in this, which was really nice. As adults that's not always something that we pursue. It was good.
3. [Coalescing] felt nice, yeah. It felt lovely. There are points where [you wonder] 'are they playing with me?' and [I realised] 'yes, we're playing with each other'. It's a comfortable feeling.
4. I was just trying to make my own patterns and overwrite everything else, and come up with new ideas, and march through everything. That's me in a social situation.
5. With a lot of this, it was just so much fun playing with you guys. Which, I mean, I'm an adult, you sound like adults, that's not something that happens often.
6. I think there was a point where someone was drawing lines with traces and I couldn't help but draw some circles around their lines and start playing around like that. Almost like kids with sparklers, or something.
7. It was very interesting and engaging [to play with one another].
8. [Coalescing] was quite playful. That was quite fun.
9. I think it was playful to interact with people, almost not as people but as people in a different form, or maybe not as *physical* people, is what I mean. And to see these different energy balls moving around. That was different to what you could do otherwise, and fun.
10. I was inspired to dance on the top of the table at one point.
11. I enjoyed how playful it was. It reminded me how fun humans are in their capacity for play.
12. It reminded me a little of writing with fireworks, like when you had sparklers as a kid.
13. **I felt playful rather than deep.**
14. **Playful connection — a sense of love.**
15. **It was very playful, and that part was fun. It was very relaxing.**
16. **Felt very playful as we continued.**
17. **The changing mudra light was fun.**
18. *Playful exploration of myself in space and flowing into connections with others.*
19. *A feeling of being grounded, a playfulness facilitated in a calm way.*
20. *That there were a small number of participants, so we could interact in a small group. Everyone was friendly and playful.*
21. *The creative aspect. Drawing lines. Pressing my fingers to create energy.*
22. *It was very fun, very unique. I feel privileged to have had the opportunity to take part.*
23. *It was fun.*
24. ***I'm not a big gamer and hesitate to make a generalized comparison to multiplayer interactive games. While mutually interactive and peaceful, a similar experience might be reached through guided meditation together while holding hands with strangers with your eyes closed. The visuals were cool, and there was a degree of intimacy, and the technology was not 'in the way' but rather helping to connect in different locations, which is certainly cool. Overall, I thought it felt more like playing an interactive game with fun teammates than a spiritual experience.***

**Theme: Transcendence of Space and Time**

1. It was a funny feeling when the thing went through the floor. It felt like the space opened up to me.  I'd been aware of the confines of [the space], vaguely.  And then we pushed it through the floor and it was very very odd to feel like everything opened out.
2. [The molecule] was just so going down beneath the floor, I couldn't comprehend it.
3. I had no idea who anybody was...I have no idea where I am in the space so I'm quite interested, when I open my eyes, where on Earth I'll be.
4. [The molecule] going into the ground, that's just so fascinating that you can overcome the physical boundaries of our reality, that always blows my mind.
5. I think I was imagining things didn't have the same human scale, so at times it felt like this could be a ginormous space and we were a completely different scale, or it could be a microscopic space, and it didn't really matter, actually.
6. I've lost perception of time.
7. **It was the beginning.  I didn't feel like there was an end.  Actually thought I was in space.**
8. **I couldn't believe how much my perspective changed when we put the molecule into the ground.**
9. **It seemed divorced from space and time.  I had a hard time giving in to the concept without trying to analyze what was going on.  Very enjoyable experience.**
10. **I didn't feel a sense of time.  I was surprised at the end that it had been an hour.**
11. *I liked being unaware of restrictions like time.*
12. *Looking down and seeing the world under the floor.*
13. *I liked moving the object and unravelling and when we put it through the floor.  I liked the parts that allowed me to view the world in a way that isn't possible in real life.*

**Theme: Noetic Quality**

1. I got quite emotional, for a moment. And then the experience was guiding us forward, so it was fleeting. But I got this surge of emotion where I don't know if I wanted to gasp or cry or what it was, but I was kind of shocked in awe.
2. That's definitely a spiritual experience of some sort. It's, like, tangible.
3. I didn't feel any nervousness. Seeing all the lights together reminded me of a dream I had years ago where the universe was ending and everybody was dissolving into balls of light and merging into one. So, I was like 'oh cool! The universe is ending. I'm cool with that'.
4. **Like awakening from a most blissful dream, that leaves its loving traces in body, heart and mind. A remembrance, a return to the simplicity and purity of being, that which we always are and can at time forget. Thank you for the remembrance.**
5. **The power and light in each individual — how this can be more potent with guidance and moving together.**
6. **Having everyone be portrayed as balls of light seemed like a very true representation of a person. Just sort of struck home as a true way to see someone.**
7. **I enjoyed intermingling lights with everyone. Got a little teary-eyed during that part, I miss people so much with COVID. And making those connections felt quite profound.**
8. *Pure unity.*
9. *I really did gather a DMT/death like Becoming for my reality. The experience felt fun, fleeting, VERY special, and at times holy. I would not say that in all my teachings and explorations that this ceremony dissolved ego in the sense that one can achieve in Healing and with inward work. This experience felt galactic, MUCH bigger in terms of the sense of Soulfulness, meaning after death, then what?*
10. *The abstraction and the time given to becoming connected to that abstraction.*
11. *The distillation of existence.*

**Theme: Comparison to Other Altered States**

1. [Taking psychedelics] wasn't quite like this, but I guess there's elements of it, for sure.
2. **Similar to experiences that I have had as somatic visions through medicine plants. The interconnective nature of energy/intention and the 'strings' that appear to interconnect us all with all living matter. This also related to childhood dreams I had prior to any 'psychedelic experience'.**
3. **About 5 days ago I tried mushrooms for the first time – if I hadn't have done that I could have selected 5 for most of them (most I've ever felt lost sense of time/space/ego).**
4. **It was like psychedelics without the psychedelics.**
5. **I wonder [how] people with and without psychedelic experiences relate to Isness.**
6. **There was something common with psychedelic experience.**
7. *I experienced the sense of interconnectedness that I would I've only previously noticed with the help of psychedelics in the right setting.*
8. *The experience brought up memories for me of medicine plant journeys and realisations that have shaped my life and what give my life meaning. These realisations are very much connected with going beyond form and finery conceptions, and also beyond mainstream conceptions of leading a good life.*
9. *Felt psychedelic.*
10. **I wondered before and during if one's prior experience with meditation and psychedelics significantly impacts how one experiences Isness. I've got some experience with both and didn't feel 'blown away' by the ceremony, although I did enjoy it and feel it has potential. I did not feel I was being critical or judgmental during it, I was simply experiencing it. I did feel more peaceful / less agitated and the end and connected to the others. Everyone was kind, sweet, and playful.**

**Theme: Sense of Beauty**

1. When I let go of needing to have control over [the molecule], it actually became spectacularly beautiful. And, actually, sitting motionless and watching what else was going on. And there was a bit, I think, where [participant's names] were dancing and interacting and it was so beautiful. So yeah, there was something about letting go of control.
2. I was amazed [by] how moved I was. I think it was the juxtaposition of the frustration and then the beholding beauty. I think the music really helped. It felt very beautiful to sit back and witness. I don't normally get moved quite so much.
3. [Molecules] are so beautiful.
4. It felt quite powerful and quite beautiful, really. The idea of interacting with abstract entities but were at the same time real entities, just not physically in this space, yet at the same time visible, and touchable, and present.
5. Yes, thank you. I was a little bit doubtful in the beginning, I didn't know what to expect and what was going to happen. It was a beautiful experience.
6. ...the whole sort of connectedness of it. The sort of weird amorphous space that you end up taking up and then just the mingling with everybody and the absence of borders and things like that...was just a very very beautiful experience.
7. **Respect, awe, love for the energy being — felt its innocence! And beauty.**
8. **The connector object is beautiful when it opens but seems too flickering when in the middle. Reminds me of flash [illegible] when I have a migraine.**
9. *Very beautiful visuals, audio and emotions that came with it!*

**Theme: Reflection on Death or Loss**

1. It sounds really cheesy, but it's given me a deep understanding of maybe what happens after life ends. I feel a bit more connected to those people that have gone, having now seen this.
2. It's clear that the energy isn't just gone, it goes somewhere. It's beautiful to think that it's all out there somewhere and it's still circling you constantly.
3. [I feel] sad that [the lights are] going away. [I feel] a lot more relaxed than I felt when I came in.
4. [When the lights faded] Does this mean we're dead?
5. There was one moment, we were talking about seeing past energies and I was thinking about my cat, whom I lost last year. I really miss her. I could really see her and the [?] was writing her name. It was beautiful, thank you.
6. **That was super cool! I arrived feeling pretty anxious and disconnected from others, but after that experience I feel much more calm and hopeful. Hugely due to the focus of how energy doesn't truly disappear. I guess the fear of dying (the existential threat) is always on our minds but that experience quietened the fear. AMAZING!**
7. **On the occasions where thing went dark, I felt a little triggered. Like, OMG, everyone's gone! Or, like my senses disappeared and so I couldn't perceive anyone. Which was lonely! Noticed some agitation due to [Participant Name] having glitches, but then I was equating it to the idea of Boudicca. Because being light being felt akin to the idea of enlightenment, I thought how if anyone gets left behind it's going to leave lingering unhappiness for everyone. At one point, the energy web circle broke on [Participant Name]'s side. I wanted to mend it but couldn't seem to. Made me a little sad – for [Participant Name] and for all of us.**
8. **When meditating upon the traces of energy remaining around us (fave bit!) I felt, again, the realization of time being not linear; I felt like I was summoning a Patronus from my lost loved ones. I think this would be amazing for grief counselling.**

# Isness pre-ceremony questionnaire

**Start of Block: Introduction**

Q23 On the following pages are a few quick questions for you to complete before starting the Isness virtual reality ceremony. Please read the questions carefully.

Thank you!

**End of Block: Introduction**

**Start of Block: Unique code**

Q26 Please enter a unique code / text (e.g. yellowsubmarine123). You will need to enter the same information on a second form after Isness has finished - this is so we can link your data and keep you anonymous.

________________________________________________________________

**End of Block: Unique code**

**Start of Block: Demographics**

Q28 What country are you participating from?

________________________________________________________________

Q27 What time is your Isness session due to start? Please write as time and your time zone (e.g. 5pm GMT)

________________________________________________________________

Q29 What is today's date? (format required is DD/MM/YYYY)

_______________________________________________________

**End of Block: Demographics**

**Start of Block: Inclusion of Community in the Self (ICS) scale.**

Q14 Below are a series of circles. The circle on the left of the pair represents you, while the circle on the right represents the other people who will be participating in your virtual reality Isness experience (who you've not yet met).

Which picture best describes your relationship with the other Isness participants?

○ a (4)

○ b (5)

○ c (6)

○ d (7)

○ e (8)

○ f (9)

**End of Block: Inclusion of Community in the Self (ICS).**

**Start of Block: Openness to the upcoming experience scale**

Q17 Please rate to what extent the statements below apply to your feeling right now, **at this present moment.**

On the scale below each statement, mark with an "X" a number between 0 and 100, where 0 = Strongly disagree and 100 = Strongly agree. There are no right or wrong answers. Do not spend too much time on any one statement but give the answer which seems to best describe your present feelings. This is a survey which is used for personal development ceremonies, so some of the questions may not be too relevant for the virtual reality experience you are about to have.

                              Strongly disagree       Strongly agree

                              0   10  20  30  40  50  60  70  80  90  100

| | |
|---|---|
| I am preoccupied with my work and/or life duties. () | |
| I feel open to the upcoming experience. () | |
| I feel well prepared for the upcoming experience. () | |
| I have a clear intention for the upcoming experience. () | |
| I feel comfortable about the upcoming experience. () | |
| I have strong expectations for the upcoming experience. () | |
| I am in a good mood. () | |
| I feel anxious. () | |
| The environment/setting feels good for my upcoming experience. () | |
| I have a good feeling about my relationship with the group/people who will be with me during my experience. () | |
| I have a good relationship with the main person/people who will look after me during the upcoming experience. () | |
| I feel able to receive without fighting back. () | |
| I have a lot of things I should be doing right now. () | |
| I have a lot on my mind right now. () | |
| I feel I have endless time to devote to this experience. () | |

Page Break

Q9 Thank you for your responses - Isness will start soon.

End of Block Openness to the upcoming experience scale

# Isness post-ceremony questionnaire

Q26 Thank you for taking part in Isness. There are now some further questions for you to answer about your experience.

------------------------------------------------------------

Q27 Please enter the unique code / text you created in the pre-Isness survey (e.g. yellowsubmarine123).

________________________________________________________________

Q18

The following questions are about your memory of the Isness ceremony experience. Try to think of the most intense part, the peak of the experience.

                              Not at all         Very much so

                  0  10  20  30  40  50  60  70  80  90  100

| | |
|---|---|
| 1. I remember my emotional experience clearly, and with confidence could describe how I felt. () | 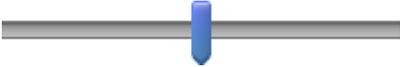 |
| 2. I remember my surroundings exactly and with confidence can recall what the environment looked like. () | 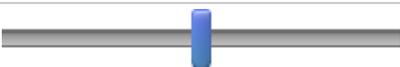 |
| 3. The peak experience exists in my mind like a very clear photograph or video. () | 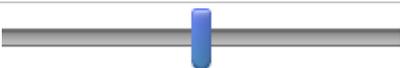 |

**End of Block: (ii) Peak experiences questionnaire**

**Start of Block: (iv) Communitas scale**

Q19

Please indicate how strongly you agree or disagree with the following statements about your Isness ceremony experience, taking into account that 1=Strongly disagree and 7= Strongly agree

| Strongly disagree | | Neither agree nor disagree | | Strongly agree |
|---|---|---|---|---|
| 1    2 | 3 | 4 | 5 | 6    7 |

| | |
|---|---|
| During the ceremony, I felt a bond with my fellow participants that I could not experience outside the ceremony. () | |
| During the ceremony, I felt a sense of belonging with the other participants. () | |
| During the ceremony, I felt a sense of harmony with the others. () | |
| During the ceremony, I felt a sense of sharing with the other participants. () | |
| The ceremony really allowed me to get to know the other participants. () | |
| During the ceremony, I felt that social status became irrelevant. () | |
| During the ceremony, I experienced all participants as equal based on basic shared humanity. () | |
| During the ceremony, ego-related tensions between participants seemed to resolve. () | |
| During the ceremony, I felt a strong connection to another participant. () | |
| During the ceremony, I felt a strong connection the facilitator. () | |

**End of Block: (iv) Communitas scale**

**Start of Block: (v) MEQ 30**

Q27

For each of the following, please rate the extent to which you agree with each statement. Looking back on the entirety of your ceremony experience, please rate the degree to which at any time you experienced the following phenomena. Answer each question according to your feelings, thoughts, and experiences at the time of the ceremony.

| None, not at all | So slight cannot decide | Slight | Moderate | Strong (equivalent in degree to any other strong experience) | Extreme (more than any other time in my life and stronger than 4) |
| --- | --- | --- | --- | --- | --- |
| 0 | 1 | 2 | 3 | 4 | 5 |

| | |
|---|---|
| 1. Loss of your usual sense of time. () | 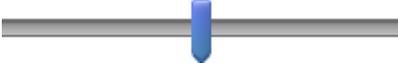 |
| 2. Experience of amazement. () | 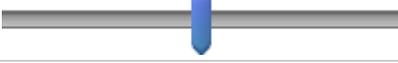 |
| 3. Sense that the experience cannot be described adequately in words. () | 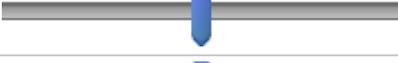 |
| 4. Gain of insightful knowledge experienced at an intuitive level. () | 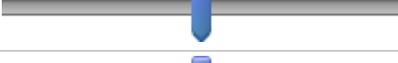 |
| 5. Feeling that you experienced eternity or infinity. () | 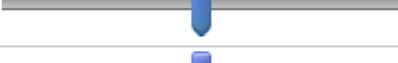 |
| 6. Experience of oneness or unity with objects and/or persons perceived in your surroundings. () | 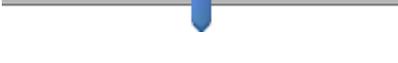 |
| 7. Loss of your usual sense of space. () | 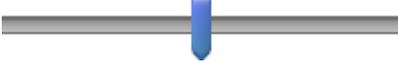 |
| 8. Feelings of tenderness and gentleness. () | 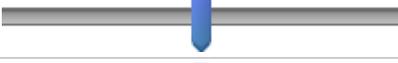 |
| 9. Certainty of encounter with ultimate reality (in the sense of being able to "know" and "see" what is really real at some point during your experience. () | 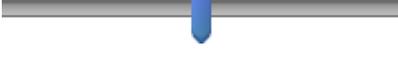 |
| 10. Feeling that you could not do justice to your experience by describing it in words. () | 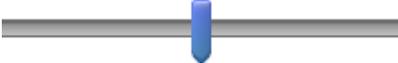 |
| 11. Loss of usual awareness of where you were. () | 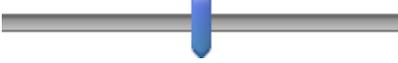 |
| 12. Feelings of peace and tranquility. () | 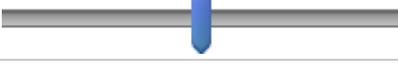 |
| 13. Sense of being "outside of" time, beyond past and future. () | 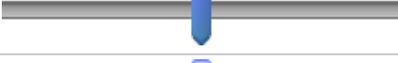 |
| 14. Freedom from the limitations of your personal self and feeling a unity or bond with what was felt to be greater than your personal self. () | 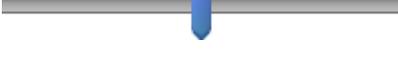 |
| 15. Sense of being at a spiritual height. () | 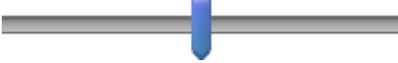 |
| 16. Experience of pure being and pure awareness (beyond the world of sense impressions). () | 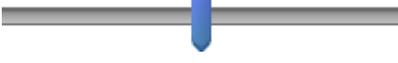 |
| 17. Experience of ecstasy. () | |
| 18. Experience of the insight that "all is One". () | |
| 19. Being in a realm with no space boundaries. () | |
| 20. Experience of oneness in relation to an "inner world" within. () | |
| 21. Sense of reverence. () | |
| 22. Experience of timelessness. () | |
| 23. You are convinced now, as you look back on your experience, that in it you encountered ultimate reality (i.e., that you "knew" and "saw" what was really real). () | |
| 24. Feeling that you experienced something profoundly sacred and holy. () | |
| 25. Awareness of the life or living presence in all things. () | |
| 26. Experience of the fusion of your personal self into a larger whole () | |
| 27. Sense of awe or awesomeness. () | |
| 28. Experience of unity with ultimate reality. () | |
| 29. Feeling that it would be difficult to communicate your own experience to others who have not had similar experiences. () | |
| 30. Feelings of joy. () | |

**End of Block: (v) MEQ 30**

**Start of Block: (vi) Ego-dissolution inventory (EDI) (Nour et al., 2016)**

Q20

Please rate to what extent the following statements apply to your Isness ceremony experience by choosing a number between 0 and 100 on the scale below each statement. Note that zero corresponds to your normal waking consciousness.

|  No, not more than usually | | | | | | | | | | Yes, I experienced this completely |
|---|---|---|---|---|---|---|---|---|---|---|
| 0 | 10 | 20 | 30 | 40 | 50 | 60 | 70 | 80 | 90 | 100 |

| Statement | |
|---|---|
| I felt especially assertive. () | |
| I experienced a dissolution of my "self" or ego. () | |
| I felt more important or special than others. () | |
| I felt at one with the universe. () | |
| My ego felt inflated. () | |
| I felt a sense of union with others. () | |
| I felt especially sure-of-myself. () | |
| I experienced a decrease in my sense of self-importance. () | |
| I felt especially keen and competitive. () | |
| I experienced a disintegration of my "self" or ego. () | |
| I felt my viewpoint was worth more than other peoples' () | |
| I felt far less absorbed by my own issues and concerns. () | |
| I felt especially self-confident. () | |
| I lost all sense of ego. () | |
| I felt especially self-assured. () | |
| All notion of self and identity dissolved away. () | |

**End of Block: (vi) Ego-Dissolution Inventory (EDI) (Nour et al., 2016)**

**Start of Block: (x) Inclusion of Community in the Self (ICS)**

Q37 Below are a series of circles. The circle on the left of the pair represents you, while the circle on the right represents the other people participating in your virtual reality Isness experience

Which picture best describes your relationship with the other Isness participants?

○ a (4)

○ b (5)

○ c (6)

○ d (7)

○ e (8)

○ f (9)

**End of Block: (x) Inclusion of Community in the Self (ICS)**

**Start of Block: (xii)**

Q21

Please answer according to your feelings towards your Isness ceremony group, taking into account that 1=Not at all and 7= All of the time

|  | Not at all | Almost never | Occasionally | Sometimes | Frequently | Almost always | All of the time |
|---|---|---|---|---|---|---|---|
|  | 1 | 2 | 3 | 4 | 5 | 6 | 7 |

| | |
|---|---|
| I felt that we were one () | 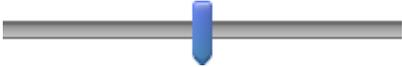 |
| I felt more sensitive to my emotions and feelings because I was surrounded by people who felt the same. () | 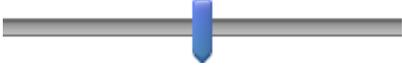 |
| I felt a strong shared emotion. () | 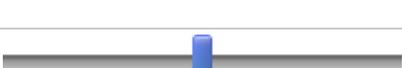 |
| I felt really united with the group, almost melded into one. () | 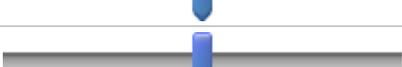 |
| I felt more intense emotions because we all went through the same experience. () | 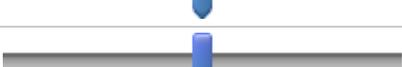 |

**End of Block: (xii)**

**Start of Block: (xiii) Bodily effects**

Q22

Have you experienced any of these bodily effects during the ceremony? Tick all that apply.

- ☐ Nausea or vomiting (1)
- ☐ Diarrhea (2)
- ☐ Body aches (3)
- ☐ Ringing in the ears (4)
- ☐ Headache (5)
- ☐ Sweats or chills (6)
- ☐ Trembling (7)
- ☐ Crying (8)

**End of Block: (xiii) Bodily effects**

**Start of Block: Demographics**

Q11 Who is your host for the experience?

○ Vadim  (4)

○ Dave  (5)

○ Rhos  (6)

○ Olivia  (7)

○ Mike  (8)

○ Greg  (9)

○ Justin  (10)

○ Harrison  (11)

○ Jeremy  (12)

○ Matthew  (13)

○ Jamie  (14)

○ Rachel  (15)

○ Sara  (16)

○ Joe  (17)

---

Page Break

Q13 What is your age in years?

_________________________________________________

Q13 What is your gender?

○ Male  (1)

○ Female  (2)

○ Other  (4)

○ Decline to state  (5)

Page Break

Q12 How many times have you had a virtual reality experience in the past 12 months?

_________________________________________________

Q25 Have you taken part in Isness before?

○ Yes  (1)

○ No  (2)

**End of Block: Demographics**

**Start of Block: Final comments**

Q29 Were there any technical problems that you noticed? If so, how did these impact the experience?

___________________________________________________

___________________________________________________

___________________________________________________

___________________________________________________

___________________________________________________

Q30 What did you *like* about the experience?

___________________________________________________

___________________________________________________

___________________________________________________

___________________________________________________

___________________________________________________

Q31 What did you *dislike* about the experience?

___________________________________________________

___________________________________________________

___________________________________________________

___________________________________________________

___________________________________________________

Q32 How do you think the experience could be *improved*?

______________________________________________________

______________________________________________________

______________________________________________________

______________________________________________________

______________________________________________________

Q33 Yes - the whole time  (1)

○ Yes - some of the time  (4)

○ No  (2)

○ Unsure  (3)

Q34 How, if at all, did their presence / absence impact your experience?

______________________________________________________

______________________________________________________

______________________________________________________

______________________________________________________

______________________________________________________

Q28 Do you have any other comments about the experience?

If you would prefer, your host will have paper and pen (please write your unique code at the top of the paper so we can link your data).

_______________________________________________

_______________________________________________

_______________________________________________

_______________________________________________

_______________________________________________

**End of Block: Final comments**

**Start of Block: Debrief**

Q10 **Debrief**

 Thank you for taking part in Isness.

 Isness is a multi-person virtual reality experience that we have created. Here we are investigating how running Isness in a distributed (across multiple sites) manner impacts the Isness experience. We are particularly interested in the impact of Isness on feelings of connectedness between participants.

 If you have any further questions, please do not hesitate to get in touch with the research team
 Dave Glowacki (glowacki@bristol.ac.uk) or Olivia Maynard (Olivia.maynard@bristol.ac.uk).

If you would like the study team to contact you with details of this study or other work they're doing, you can enter your email address at the following link - this separates your email address from your study data, keeping your responses anonymous

https://bristolexppsych.eu.qualtrics.com/jfe/form/SV_6mSi4TBYy0Ir9CR

End of Block: Debrief